\documentstyle[mprocl,feynmf]{article}
\newcommand{\LQCD}[0]{\Lambda_{\rm QCD}}
\newcommand{\darr}[1]{\raise1.8ex\hbox{$\leftrightarrow$}\mkern-17.0mu #1}
\newcommand{\ie}[0]{{\em ie}}
\newcommand{\onec}[0]{\hbox{$\frac1c$}}
\newcounter{myfigcount}
\newsavebox{\myfignum}
\newenvironment{myfig}[1]%
{\refstepcounter{myfigcount}%
\savebox{\myfignum}{\parbox{4.0in}{{\vspace*{0.6cm}\small 
Fig.~\arabic{myfigcount}.\ #1}}}%
\vspace{1.0cm}\begin{center}}%
{\usebox{\myfignum}\end{center}\vspace{1cm}}
\input{psfig}
\def\Journal#1#2#3#4{{#1} {\bf #2}, #3 (#4)}

\def\NPB{{\em Nucl. Phys.} B}
\def\PLB{{\em Phys. Lett..}  B}

\def\PRD{{\em Phys. Rev.} D}
\def\ZPC{{\em Z. Phys.} C}
\begin{document}
\begin{fmffile}{paperfmf}
\title{A MODERN INTRODUCTION TO QUARKONIUM THEORY}
\author{ BENJAMIN GRINSTEIN}
\address{Department Of Physics, University of California, San Diego\\
La Jolla, CA 92093-0319, USA\\E-mail: bgrinstein@ucsd.edu}

\maketitle \abstracts{ Recent advances in lattice and continuum QCD
have given us new insights into quarkonium physics. These set of
lectures are intend for the uninitiated. We first give a physical
picture of quarkonium and describe the hybrids states established in
lattice QCD. Then we give an unorthodox presentation of
Non-Relativistic QCD (NRQCD) including a novel method for the
application of spin-symmetries. Finally we describe the prototypical
application of NRQCD: cancellation of infrared divergences in decays of
P-wave quarkonia.}
%
\section{Introduction}
\subsection{About these lectures}
I have put together a little material on the theory of quarkonia that
I hope will be a good introduction to the subject for novices, be them
graduate students or researchers outside this field. I must warn the
readers that I give a personal view of the subject. There are many
excellent, more orthodox reviews available. However, I have combined
material in a way that I think brings out the physics more
clearly. For example, although my aim is to present the advances in
NRQCD of the last few years, at the expense of the whole subject of
quarkonium production I have used much space and effort in giving a
physical picture of quarkonium and hybrids based on the rather good
recent calculations of inter-quark potentials in lattice QCD. 
I also have injected here and there little tricks that help me
understand or calculate. For example, I have given explicitly a
foolproof method of calculating spin-symmetry relations using the
Wigner-Eckart theorem by means of a trace formula, akin to that used
in HQET. 

There are three parts to these lectures. First we discuss the physical
picture. To this end we briefly use the bag-model picture to motivate
a non-trivial inter-quark potential, linear  at long distances. Then
the role of octets in the description creeps in through the discussion
of hybrids. Hybrids are the particle physics analogs of molecules, so
it should be no surprise that a Born-Oppenheimer approximation can be
used to describe them (the BO is reviewed). This immediately suggests
we study QCD in the Non-Relativistic limit, or `NRQCD', which is the
subject of the middle part of these lectures. The last part describes
the application of NRQCD to the decays of S and P-wave quarkonia. 

Since these lectures are intended for learning students, I have toiled
to include many exercises. They are included as separate paragraphs,
and most often the answer is given.

I have left many loose ends untied. This is not only because I have been
careless. There are many questions that are unresolved, and I hope the
alert reader will spot them and hopefully solve them. In the
conclusions I attempt to list some of these questions, but offer
little in the way of solutions.

\vspace{1cm}
\begin{myfig}{Static quarks at small separation. The shaded region
represents the non-perturbative QCD vacuum. The `bag' of perturbative
vacuum has radius $R_b$ given in Eq.~(\ref{eq:bagradius})}
\label{fig:QCDbag}
\hfil\psfig{figure=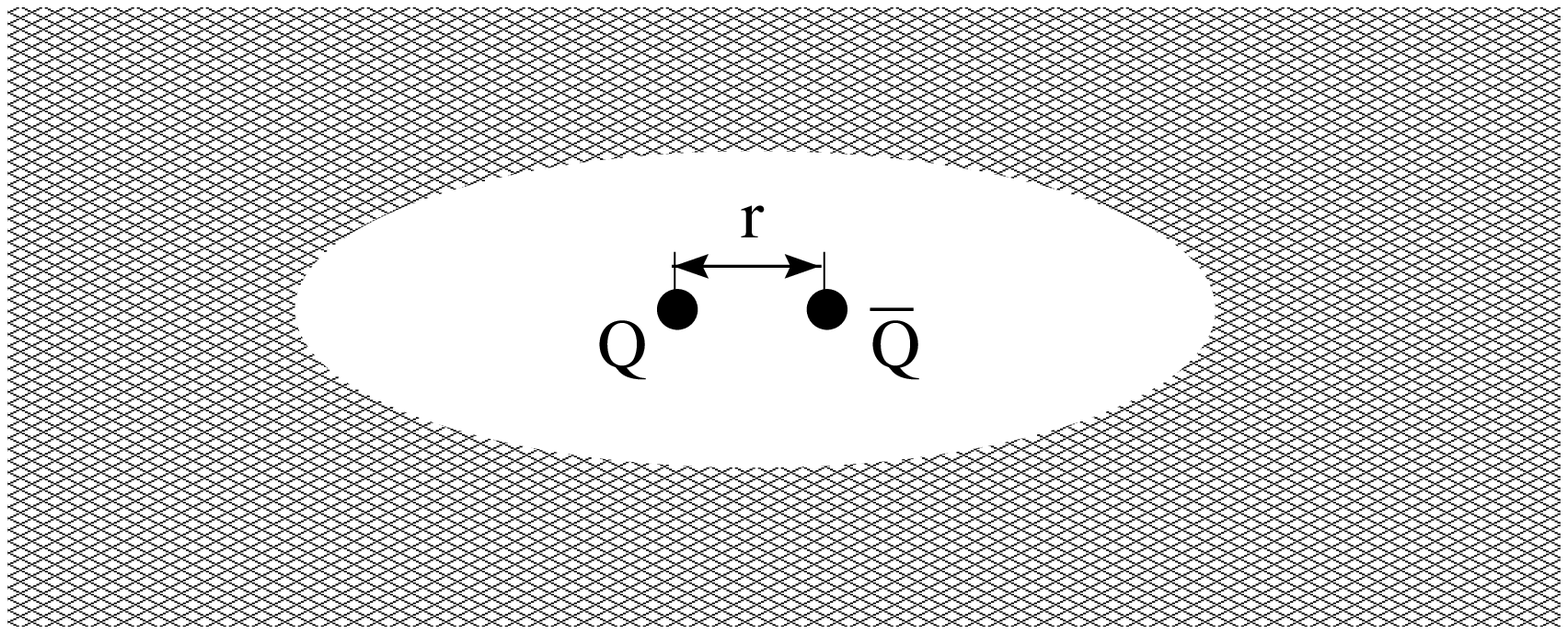,height=1.5in,width=4.0in}\hfil
\end{myfig}

\subsection{Quarkonium: A Physical Picture}
Consider a quark $Q$ and an antiquark $\bar Q$ pinned down a distance
$r$ from each other. Let us begin our considerations by assuming that
this distance $r$ is much smaller than the typical size of a hadron
$\LQCD^{-1}\sim1$~Fermi. See Fig.~\ref{fig:QCDbag}. Far away from the
pair, at a distance $R\gg r$, the chromo-electric field is that of a
dipole,
\begin{equation}
E_\theta\sim g(r)\frac{r\cos\theta}{R^3},
\end{equation}
where $g(r)$ is the coupling constant of QCD at distance scales of
order $r$. This description in terms of a perturbative vacuum cannot
be correct at very large distances from the dipole. When the field
strength drops to a critical value \(E_{\theta,\rm crit}\sim \LQCD^2\)
the vacuum quickly turns non-perturbative. Thus the dipole can be
described as a perturbative ``bag'' inside the non-perturbative
vacuum.\cite{HHKR,Ono,JKM2} The radius $R_b$ of this bag is thus determined
\begin{equation}
\label{eq:bagradius}
\frac{g(r)}{R_b^3}\sim\LQCD^2.
\end{equation}

\begin{myfig}{At short distances the inter-quark force is
Coulomb-like, and arises from this one gluon exchange diagram.}
\label{onegluonexch}
\begin{fmfgraph}(60,20)
\fmfleft{i,j}
\fmfright{k,l}
\fmf{plain_arrow}{i,x,k}
\fmf{plain_arrow}{l,y,j}
\fmf{curly,tension=0}{x,y}
\end{fmfgraph}
\end{myfig}

We will be interested in dynamical bound states, with quarks free to
move. For this we need an understanding of the force between static
quarks as a function of distance $r$. For $r\ll R_b$ the interaction
is well approximated as a Coulomb force. It arises from perturbative
single gluon exchange; see Fig.~\ref{onegluonexch}. The energy of the
state is the energy of the Coulomb dipole, $\sim1/r$. At the opposite
extreme, shown in Fig.~\ref{fig:QCDbagsquashed}, when $r\gg R_b$
non-perturbative effects become important. The perturbative vacuum is
squashed out into a cigar shape; this is the ``string limit''.  The
bag carries volume energy density ${\cal E}$. If the cross sectional
area of the cigar in the string limit is $A$, the energy of the state
is ${\cal E}\cdot (Ar)$, \ie, it increases linearly with $r$.  The
linear slope of this potential energy is the ``string tension'' ${\cal
E}A$.\footnote{In fact the cross sectional area of the string solution
is not an independent parameter, $A\sim\sqrt{g^2/{\cal E}}$, so the
string tension $\sim\sqrt{g^2{\cal E}}$.}

\vspace{1cm}
\begin{myfig}{For large inter-quark separation $r$ the perturbative
vacuum `bag' takes on a cigar shape, of fixed cross sectional area
$A$. The energy of the configuration is proportional to the volume of
the bag, $Ar$, giving rise to a linear inter-quark potential.}
\label{fig:QCDbagsquashed}
\hfil\psfig{figure=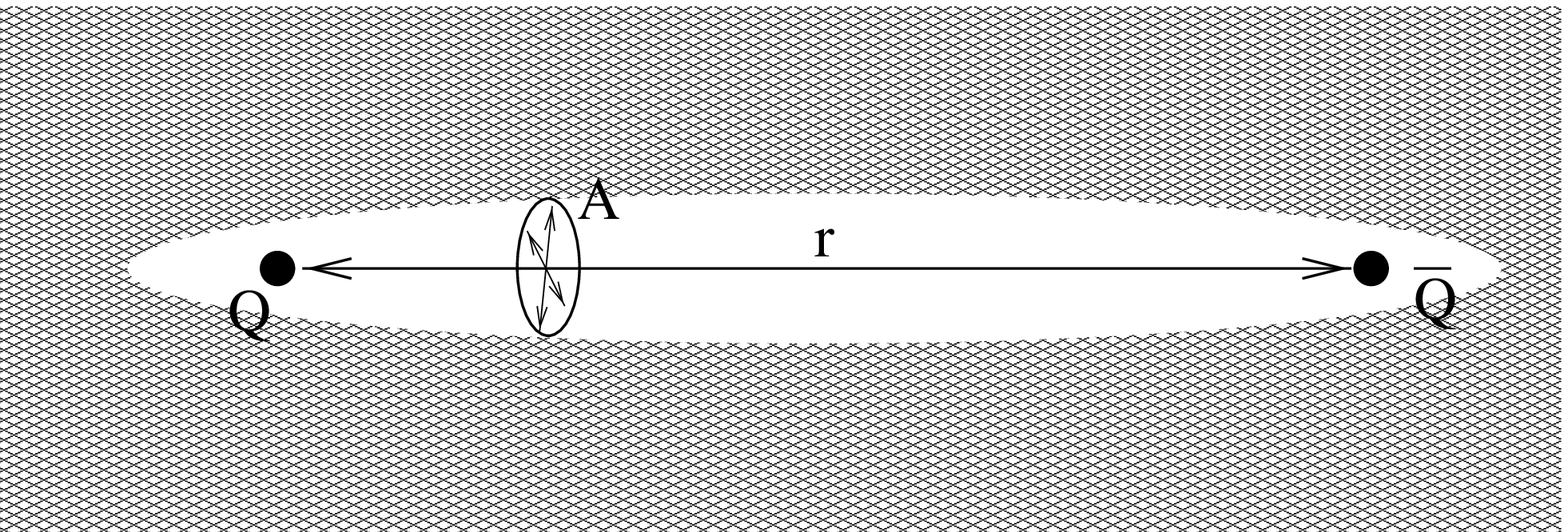,height=1.5in,width=5.0in}\hfil
\end{myfig}

\newpage

\subsection{Digression: The strength of the Coulomb Gluon}
In the perturbative vacuum the inter-quark force is dominated by
single gluon exchange, as in the following figure:

\begin{myfig}{The gluon exchange gives rise to a Coulomb interaction.} 
\label{eq:coulomb}
\[
\parbox{60mm}{\begin{fmfgraph*}(60,20) \fmfleft{j,i} \fmfright{l,k}
\fmfv{label=$j,,\vec p$}{i} \fmfv{label=$l,,-\vec p$}{j}
\fmfv{label=$i,,\vec p\hskip2pt{}'=\vec p+\vec q$}{k} \fmfv{label=$k,,-\vec
p\hskip2pt{}'$}{l} \fmf{plain_arrow}{i,x,k} \fmf{plain_arrow}{l,y,j}
\fmf{curly,tension=0,label=$q$,label.dist=60}{x,y}
\end{fmfgraph*}}
\qquad\qquad\sim \frac1{{\vec q}\,^2} \sum_a T^a_{ij}T^a_{lk}
\]
\end{myfig}

\noindent
The momentum dependence gives rise to a Coulomb potential (the Fourier
transform of $1/{\vec q}\hskip2pt{}^2$ is proportional to $1/r$). Here
we will focus on the color factor, \(\sum_a T^a_{ij}T^a_{lk}\). The
matrices $T^a$ are the generators of $SU(3)$ in the fundamental
representation, that is, Gell-Mann matrices. The indices $i$, $j$,
$k$, $l$ refer to the external quarks, as in Fig.~\ref{eq:coulomb}.

It is now easy to see that the strength of the interaction depends on
the relative color of the $Q\bar Q$ state. Since  $Q$ is in the
$\bf 3$ (fundamental representation) of $SU(3)$, and 
\[{\bf 3}\times{\bf\overline3}={\bf1}\oplus{\bf8},
\]
the $Q\bar Q$ force is in either of these two channels. It will be
instructive to carry the calculation for the more general case of
$SU(N)$, so the representation $\bf8$ is understood as the adjoint
representation of SU(N). To find the
\({\bf1}\) and \({\bf8}\) components of the force we prepare the
initial $Q\bar Q$ state to be purely \({\bf1}\) or~\({\bf8}\):
\begin{eqnarray}
(Q_j\bar Q_l)_1 & \equiv& \delta_{jl} (Q\bar Q)\\
(Q_j\bar Q_l)_8 & \equiv& Q_j\bar Q_l-\frac1N\delta_{jl} (Q\bar Q).
\end{eqnarray}
Contracting this with
\[\sum_a T^a_{ij}T^a_{lk}=
\frac12(\delta_{ik}\delta_{jl}-\frac1N\delta_{ij}\delta_{kl})\]
we get 
\begin{eqnarray}
(Q_j\bar Q_l)_1\sum_a T^a_{ij}T^a_{lk} & =&\frac{N^2-1}{2N}(Q_i\bar Q_k)_1\\
(Q_j\bar Q_l)_8 \sum_a T^a_{ij}T^a_{lk} & =&-\frac{1}{2N}(Q_i\bar Q_k)_8.
\end{eqnarray}

We see that the force in the color octet channel is of opposite sign
(in fact, repulsive) and weaker than in the singlet channel by a
factor of $1/(N^2-1)$.

In atomic physics the $\vec L\cdot\vec S$ interaction is computed most
easily via the well known formula
\[\langle2\vec L\cdot\vec S\rangle=j(j+1)-l(l+1)-s(s+1).\]
The  reader may verify the analogous formula
\[\langle \sum_a T^aT^a\rangle_R=C(R)-2C({\rm fund})\]
where $C(R)$ is the Casimir invariant of the representation $R$,
\[\sum_a T^aT^a=C(R)\bf1.\]
For $SU(3)$ if we normalize $C(R)$ so that $C({\bf 8})=3$, then
$C(R)=4/3,10/3$ and~6, for $R={\bf 3}, {\bf6}$ and $\bf10$,
respectively. 

\begin{myfig}{One gluon exchange gives rise to gluon-quark force. The
force is attractive in the {\bf 3} and {\bf6} channels, but repulsive
in the {\bf15} channel.}
\label{fig:gqforce}
\begin{fmfgraph}(60,20)
\fmfleft{j,i}
\fmfright{l,k}
\fmf{quark}{i,x,k}
\fmf{gluon}{j,y,l}
\fmf{curly,tension=0}{x,y}
\end{fmfgraph}
\end{myfig}

One can compute the force between a gluon and a quark similarly; see
Fig.~\ref{fig:gqforce}. The available channels are given by
\[{\bf 3}\times{\bf8}={\bf15}\oplus{\bf\overline6}\oplus{\bf3}.
\]
The force is attractive in the $\bf3$ channel,
\[C({\bf3})-C({\bf8})-C({\bf3})=-3,
\]
and in the ${\bf\overline6}$ channel,
\[C({\bf\overline6})-C({\bf8})-C({\bf3})=-1,
\]
but repulsive in the $\bf15$ channel.

\subsection{Hybrids: Octet Quarkonia}
\label{sec:hybrids}
As we have seen a quark-antiquark pair in a color octet configuration
repel each other, and a quark-gluon pair attract each other in either
the $\bf3$ or $\bf\overline6$ configurations. A state of quarks in an
octet configuration can bind to a gluon, provided the gluon attraction
to both quarks overcomes the repulsion between quarks. States like
this in which the glue state plays a particle-like role in binding are
called ``hybrids'', because they are a hybrid of a glueball and a pure
quarkonium state. Lattice calculations~\cite{JKM1} convincingly
indicate that QCD predicts such states, but experimentally no such
state has been demonstrated. The state is very non-perturbative in
nature, but can be empirically understood in terms of a bag picture in
the string limit with excited chromo-electric and magnetic cavity
modes in the bag.

Hybrids have a useful analog in atomic physics: the $H_2$
molecule. There two heavy protons repel each other, but the
intervening electronic cloud attracts them both and overcomes the
repulsion. The physics of the molecule is well described by the
Born-Oppenheimer (BO) approximation, which we now review. 

\subsection{Quick Review: Born-Oppenheimer Approximation}
Consider~\cite{Messiah} the molecular Hamiltonians 
\[ H=T_e+T_N+V \]
where the kinetic energies for electrons and nuclei are, respectively,
\[T_e=-\frac1{2m}\sum_i\frac{\partial^2}{\partial x^2_i}
\qquad,\qquad
T_N=-\sum_j\frac1{2M_j}\frac{\partial^2}{\partial X^2_j}\]
and the potential energy $V=V(X_j,x_i)$ is a function of all
particle locations, in general. 

In the BO approximation one considers first the simpler Hamiltonian
\[ H^{(0)}= T_e+V \]
obtained formally by taking the infinite nuclear mass limit. Since
$[X_j,H^{(0)}]=0$ we can simultaneously diagonalize $X_j$ and
$H^{(0)}$. Set the values of $X_j$ to $X'_j$ and notice that $X'$
appear as parameters of the Hamiltonian for the electrons. We include
a label $X'$ in
eigenstates  and eigenvalues, 
\[H^{(0)}|nX'\rangle=W_n(X')|nX'\rangle,\]
to remind us that they carry implicit dependence on $X'$. In the
$|xX\rangle$ representation,
\[
\langle xX|nX'\rangle =\phi_n(x,X')\delta(X-X')
\]
this is just the Shrodinger equation for  the electrons in the field
of fixed point charges
\[[T_e+V(x,X')]\phi_n(x,X')=W_n(X')\phi_n(x,X'). \]

Next we let the nuclei move, slowly. We replace above  $X'\to
X'(t)$. If the motion is slow enough we can solve the new Schrodinger
equation 
\[[T_e+V(x,X'(t))]\tilde\phi_n=\tilde W_n\tilde\phi_n\]
in the adiabatic approximation:~\footnote{The phase, here ignored, can
play an important role.\cite{LR,LNRZ}}
\begin{eqnarray*}
\tilde\phi_n &=& \phi_n(x,X'(t))\qquad \qquad \hbox{(up to a phase)}\cr\cr
\tilde W_n &=& W_n(X'(t))~.
\end{eqnarray*}
The condition for the validity of the adiabatic approximation is that
the probability $P_n$ of electrons jumping out of state $|nX'\rangle$
be small:
\begin{equation}
\label{eq:probofjump}
P_n=\left|\frac{\langle nX'|\left(\frac{d}{dt}|nX'\rangle\right)}%
{{\rm min}\Delta W}\right|^2\ll1,
\end{equation}
where $\Delta W$ stands for the electronic energy spacings.

To estimate $P_n$, let $a$ be the molecular size, that is, the typical
separation between nuclei. By the uncertainty principle the electronic
momentum is $p_e\sim1/a$ and thus $\Delta W\sim E_e\sim
p_e^2/m\sim1/ma^2$. For the numerator we argue as follows. To remove
one atom from the molecule we need to move that nucleus by $\delta
X'\sim a$. At this point the new state is essentially orthogonal to the
original one,
\[\int dx \phi_n(x,X'+\Delta X')\phi_n(x,X')\approx0\qquad{\rm for}\
\Delta X'\sim a\]
or, since $\phi_n$ is normalized,
\[\int dx\, a \frac{\partial\phi_n}{\partial X'}\phi_n\sim1\]
Now, denote the slow nuclear velocity by $V_j=dX'_j/dt$. We have,
roughly, 
\[\langle\phi| \frac{d\phi}{dt}\rangle
\sim\frac1{a}V_{{\rm rms}}\sim\frac1a\sqrt{\frac{t_N}{M}}\]
where the rms velocity is $V_{{\rm
rms}}=\sqrt{\sum_j\langle V_j^2\rangle}$ and the kinetic energy 
\(t_N=\sum_j\frac12M_j\langle V_j^2\rangle\). Thus
\[
P_n\sim\frac{t_N/Ma^2}{E_e(1/ma^2)}=\left(\frac{m}{M}\right)\frac{t_N}{E_e}.
\]
Now, $t_N$ has contributions from rotational and vibrational
modes. In most cases the latter is dominant, and we write $E_{{\rm
vib}}=\omega$. Since the molecule is bound by the electronic cloud,
the vibrational potential energy $V_N=\frac12M\omega^2X^2$ at $X\sim
a$ should be comparable to the electronic energy, 
\(\frac12M\omega^2a^2\sim\frac12\frac1{ma^2}\), or
\[t_N\sim E_{{\rm vib}}\sim\omega\sim\frac1{\sqrt{Mm}}\frac1{a^2}\]
Using this in $P_n$,
\begin{equation}
\label{eq:approxprob}
P_n\sim \frac{m}{M}\sqrt{\frac{m}{M}}.
\end{equation}

Exercise: Show that rotational modes give \( P_n\sim (m/M)^2\).

In the adiabatic approximation $n$ is a good quantum number (the
coupling between electronic levels is neglected). Eigenvectors of $H$
are linear combinations of $|nX'\rangle$ with $n$ fixed:
\[\int|nX'\rangle\psi(X')dX',\]
or, in the $|xX\rangle$ representation,
\[\Phi_n(x,X)=\phi_n(x,X)\psi(X)\]
The eigenfunctions of $H$ are found by expanding in terms of
these. Let's use these as trial wavefunction in the variational
method. The energy functional is 
\[E(\Phi_n)=\frac{\langle\Phi_n|H|\Phi_n\rangle}{\langle\Phi_n|\Phi_n\rangle}
= \frac{\langle\psi|H_n|\psi\rangle}{\langle\psi|\psi\rangle}\]
where 
\[ H_n\psi(X)=\int\phi_n(x,X)[H\phi_n(x,X)\psi(X)]dx, \]
and the condition $\delta E(\Phi_n)=0$ is equivalent to
\[ H_n\psi = E\psi \]
This is nothing but the Schroedinger equation for the nuclei in the
background Coulomb field of the electrons in the $n$-state.

Exercise: Show that 
\[H_n=T_N+W'_n+W_n\]
where
\[W'_n=\sum_j \frac
1{2M_j}\int\left(\frac{\partial\phi_n}{\partial X_j}\right)^2dx\]
so the nucleus moves in a potential $W_n(X)+W'_n(X)$. Show
$W'_n\sim1/Ma^2$ and is therefore a small correction.

\subsection{Born-Oppenheimer in Quarkonium and the Relevant Distance Scales}
It seems obvious that the Born-Oppenheimer approximation should be
adequate in the description of hybrids. There we would choose the
quark and antiquark to be the slow degrees of freedom while the glue
would be the fast degrees of freedom. One would guess that the
probability $P_g$ of Eq.~(\ref{eq:probofjump}), that the glue stays in
a fixed state, is approximated by Eq.~(\ref{eq:approxprob}),
$P_g\sim(\LQCD/M_Q)^{3/2}$, so for large $M_Q$ the adiabatic
approximation will be very good.

It is not so obvious that the color singlet states can also be treated
in this approximation. This is a subtle issue which I won't discuss,
save for the following guess based on atomic physics analogs. Since
the $(Q\bar Q)_1$ acts like a bound state in a mutually attractive Coulomb
field, one could guess that the average kinetic energy 
$t_{Q\bar Q}\sim1/Ma^2\sim$~``Rydberg'', where $a\sim(\alpha_sM)^{-1}$ is the
``Bohr Radius'', and if the gluonic energy (or ``mass'') is fixed
$E_g\sim m_g\sim\LQCD$,
then $P_g\sim(m_g/M)(t_{Q\bar Q}/E_g)
\sim(\LQCD/M)(\alpha_s^2M/\LQCD)\sim\alpha_s^2$. Therefore
the approximation will work to the extent that $\alpha_s^2$ is small.

The scales that appear in this discussion play an important role
throughout, so we pause to revisit them. The overall mass of the
system is dominated by $M$; the size of the system is its Bohr radius,
$a$, with $a^{-1}\sim M\alpha_s\sim Mv$; the ``Rydberg'' is the gross
measure of energy spacings and is the (inverse of the) primary time
scale in the system, $\sim Mv^2$; and the strong interactions become
relevant at distances greater than $1/\LQCD$. Just as in the Hydrogen
atom, we we have used $v\sim\alpha_s$ as the velocity of the
quarks. So we have the hierarchy of scales
\[
M\gg Mv\gg Mv^2\gg\LQCD.
\]
Whether this is realized in nature for charm or beauty is a
subject much debate, so we will develop the
theory hoping that it will be appropriate for at least one of these
systems. Notice also that theoretically one could have such  large
masses that even $Mv^n\gg\LQCD$ for some $n\ge3$. Such scales dictate
the physics of, eg,  hyperfine splittings. In such a world one would
be able to compute reliably even such small effects.

It is now clear how our discussion of the static quarks in a
perturbative bag plays a role: it just corresponds to the first step
in the BO approximation. As argued above, the inter-quark potential is
Coulombic at short distances and increases linearly with separation at
long distances. In the BO approximation this potential is used in a
Schroedinger equation for the quarks to describe the bound state.

\subsection{Towards NRQCD}
Now that we have decided to apply the Born-Oppenheimer approximation
to heavy quarkonia we find difficulties implementing this
program. Consider the Lagrangian
\[{\cal L}=\bar\psi(i\gamma\cdot D-M)\psi\]
where the covariant derivative is $D_\mu=\partial_\mu-i\frac
gcA_\mu$. Dirac's Hamiltonian density follows
\[
{\cal H}= -\bar\psi(-i\vec\gamma\cdot\vec\partial+\frac
gcA^0\gamma^0-\frac gc \vec\gamma\cdot\vec A -M)\psi.\]
We immediately face two difficulties. In the standard application of
Born-Oppenheimer we need to write a Hamiltonian that includes a
kinetic energy term for the slow degrees of freedom. Our Hamiltonian
has no obvious kinetic energy term, and it does not account separately
for the $Q$ and $\bar Q$ variables. 

The solution to this problem is to work with a non-relativistic
version of ${\cal H}$, by doing a $1/c$ expansion. This immediately
leads to what is known as ``Non-Relativistic QCD'' or NRQCD.

Now, the $1/c$ expansion in QED is old hat. You can find detailed
discussions in famous texts.\cite{BLP} In fact, the Born-Oppenheimer
approximation and the non-relativistic expansion applied to $Q\bar Q$
states in QCD is also old hat, first considered two decades
ago.\cite{NOSVVZ,Peskin,BP} It will become clear what new developments
have re-ignited interest in this field.

\section{NRQCD}
I will present NRQCD as an expansion in a parameter, namely $1/c$.
The unattractive alternative is to expand in a kinematic variable,
$v_{\rm rel}$ ---the relative velocity of the $Q\bar Q$ pair in a
slowly moving CM frame.

\subsection{Notation}
Our metric is
\[ \eta^{\mu\nu}={\rm diag}(+---). \]
The Dirac 4-spinor is broken up into two 2-spinors
\begin{equation}
\label{eq:2spinors}
\Psi=\pmatrix{\psi\cr\chi\cr}.
\end{equation}
The Dirac gamma matrices satisfy
\[\{\gamma^\mu,\gamma^\nu\}=2\eta^{\mu\nu}\]
and are given, in the Dirac representation, by
\[\gamma^0=\pmatrix{{\bf1}&0\cr0&{-\bf 1}\cr},\qquad
\gamma^i=\pmatrix{0&\sigma^i\cr-\sigma^i&0\cr},
\]
where the $2\times2$ Pauli matrices $\sigma^i$ satisfy
$\{\sigma^i,\sigma^j\}=2\delta^{ij}$ and $\sigma^3={\rm diag}(1, -1)$. 
We will also use
\[\gamma_5=i\gamma^0\gamma^1\gamma^2\gamma^3=\pmatrix{0&{\bf1}\cr{\bf1}&0}\]
and the invariant four index tensor has
\[\epsilon^{0123}=-\epsilon_{0123}=1.
\]
Finally, 
\[\sigma^{\mu\nu}=\frac i2[\gamma^\mu,\gamma^\nu]\]
is, in the Dirac representation,
\[ \sigma^{0i}=i\gamma^0\gamma^i=\pmatrix{0&i\sigma^i\cr
i\sigma^i&0\cr}
\qquad
\sigma^{ij}=-i\epsilon^{ijk}\pmatrix{\sigma^k&0\cr
0&\sigma^k\cr}
\]
\subsection{Classical NRQCD}
In the Dirac representation the two 2-spinors in Eq.~(\ref{eq:2spinors})
roughly correspond to the particle and anti-particle components of the
4-spinor. We'd like to make this explicit by rewriting the Dirac
lagrangian in terms of variables appropriate for a non-relativistic
approximation about a  frame of reference. The appropriate frame is,
of course, that for which the particles are at, or almost at, rest.
To this end shift the field so that the trivial (but dominant) time
evolution factor is taken out,
\[
\Psi=e^{-iMc^2t}\tilde\Psi,
\]
in the Dirac Equation
\[
[i\gamma\cdot D+Mc(\gamma^0-1)]\tilde\Psi =0
\]
and write out the components \(\tilde\Psi=\pmatrix{\psi\cr\chi\cr}\)
\[\pmatrix{\frac1ciD_t&-i\vec\sigma\cdot\vec D\cr
i\vec\sigma\cdot\vec D&-\frac1c(2Mc^2+iD_t)\cr}
\pmatrix{\psi\cr\chi\cr}=0.
\]
The field $\chi$ is suppressed by a power of $1/c$ so one may solve
this equation recursively. Formally, we write
\[\chi=\frac1{2Mc+\frac1ciD_t}i\vec\sigma\cdot\vec D\psi
\]
and use the field 
\[\tilde\Psi=\pmatrix{\psi\cr 
  \frac1{2Mc+\frac1ciD_t}i\vec\sigma\cdot\vec D\psi\cr}
\]
to describe the quark $Q$ being nearly on-shell,  in the rest frame of
the $Q\bar Q$ system, in the $1/c$ expansion.

Exercise: Carry out a similar expansion, starting from 
\[
\Psi=e^{+iMc^2t}\hat\Psi,
\]
and show how to describe $\bar Q$ in terms of  $\hat\Psi$ in the $1/c$
expansion. 

We can now easily compute the lagrangian that describes the quark
state,
\[
c{\cal L}=c\overline{\tilde\Psi}
[i\gamma\cdot D+Mc(\gamma^0-1)]\tilde\Psi
\]
\[
=\psi^\dagger iD_t\psi-
\psi^\dagger i\vec\sigma\cdot\vec D
\frac{c}{2Mc+\frac1ciD_t}
i\vec\sigma\cdot\vec D\psi
\]

This is not yet written as an expansion in $1/c$. To this end we
expand
\[
(2Mc+\frac1ciD_t)^{-1}=(2Mc)^{-1}(1-iD_t/2Mc^2+\cdots)
\]
and use
\begin{eqnarray*}
\sigma^i\sigma^jD_iD_j & = &
(\delta^{ij}+i\epsilon^{ijk}\sigma^k)D_iD_j\cr
& = & {\vec D}^2+\frac{g}{2c}\vec\sigma\cdot\vec B
\end{eqnarray*}
to write a $1/c$-expanded lagrangian
\[
c{\cal L}=\psi^\dagger(iD_t-\frac1{2M}(i\vec D)^2)\psi + 
\frac{g}{2Mc}\psi^\dagger\vec\sigma\cdot\vec B\psi +\cdots
\]

Exercise: The first term in the ellipsis is 
\(\vec\sigma\cdot\vec D D_t\vec\sigma\cdot\vec D\). Consider the piece
\(\vec\sigma\cdot\vec D\sigma^i[D_t,D_i]\). Show it gives
\(\vec D\cdot\vec E+i\sigma\cdot\vec D\times\vec E\). 

Exercise: Let \(A=\vec\sigma\cdot\vec D\). Write 
\(A D_t A=\frac12
A[D_t,A]-\frac12[D_t,A]+\frac12D_tA^2+\frac12A^2D_t\).
As in the previous exercise, the first two terms give no time
derivatives $D_t$ acting on $\psi$. We'd like to have time
derivatives appear in the lagrangian  only as $\psi^\dagger D_t\psi$.
Show that this can be accomplished, to this order, by redefining
\(\psi\to(1-\frac12A^2)\psi\).

Exercise: Determine the $1/c$-expanded lagrangian for the antiquark $\bar Q$.

Putting the result of these exercises together, and incorporating the
lagrangian for the fast degrees of freedom we have the following
lagrangian for NRQCD:~\cite{CL}
\begin{eqnarray}
\label{eq:lag-NRQCD}
{\cal L} & =& 
\frac12(\partial_iA^a_0-\frac1c\partial_tA_i^a-\frac{g}cf_{abc}A^b_iA^c_0)^2\cr\cr
& &-\frac14(\partial_iA^a_j-\partial_jA_i^a-\frac{g}cf_{abc}A^b_iA^c_j)^2\cr\cr
& &+\psi^\dagger(iD_t-\frac1{2M}(i\vec D)^2)\psi \cr\cr
& &+ \frac{c_Fg}{2Mc}\psi^\dagger\vec\sigma\cdot\vec B\psi \cr\cr
& &+ \frac{1}{8M^3c^2}\psi^\dagger({\vec D}^2)^2\psi\cr\cr
& &+ \frac{c_Dg}{8M^2c^2}\psi^\dagger({\vec D}\cdot\vec E
      - \vec E\cdot\vec D   )\psi\cr\cr
& &+ \frac{c_Sg}{8M^2c^2}\psi^\dagger\vec\sigma\cdot
(\vec D\times\vec E      - \vec E\times\vec D   )\psi\cr\cr
& &+{\cal O}(1/c^3)
\end{eqnarray}
Here we have rescaled $\psi$ by $\sqrt{c}$ and we have generalized the
coefficient of the fourth, sixth and seventh lines with hindsight
(although, for now, $c_F=c_D=c_S=1$). We have omitted the antiquark
terms, which are left to the student as an exercise. The terms in this
lagrangian have well known physical interpretation. The fourth line is
the chromo-magnetic moment interaction, the fifth line is the first
relativistic correction, the sixth line is the Darwin term and the
seventh is the spin orbit coupling.

It would be desirable to have a lowest order NRQCD lagrangian that
would be completely independent of $1/c$. One would then use the
$1/c=0$ spectrum as a solid basis for perturbation theory in the small
parameter $1/c$, much as it is done in the $1/M$ expansion for
$B$-mesons. If this cannot be done, that is, if our lowest order
lagrangian contains explicitly the parameter $1/c$, then the lowest
order states will depend implicitly and non-trivially on $1/c$, and an
explicit expansion in $1/c$ will be impossible. 

What are the leading terms in the 1/c expansion? To this end one is
tempted to omit from the lagrangian above all terms with powers of
$1/c$. This is almost correct except for one subtlety. The second term
on the first line of (\ref{eq:lag-NRQCD}) must be kept. Recall, for a
Hamiltonian we need the generalized momentum
\[\pi_i=\frac{\partial {\cal L}}{\partial(\partial_tA_i)}
=\frac1{c^2}(\partial_tA_i-c\partial_iA_0)
\]
so that the Hamiltonian starts at order $c^2$:
\[{\cal H}=\pi_i\partial_tA_i-{\cal L}=\frac12c^2\pi^2+\cdots\]
Also, in the third line, the covariant derivative contains hidden
dependence on $1/c$ which should be dropped. Hence, we have the leading
order lagrangian:
\[
{\cal L}_0=
\frac12(\partial_iA^a_0-\frac1c\partial_tA_i^a)^2
-\frac14(\partial_iA^a_j-\partial_jA_i^a)^2
+\psi^\dagger(iD_t+\frac{\vec \nabla^2}{2M})\psi 
\]
\subsection{Interlude: Momentum Space Derivation}
It is instructive to derive the effective lagrangian of NRQCD by
analyzing Green functions (Feynman diagrams) in momentum space. This
may be less elegant that the derivation above, but it has several
advantages. First, it is more intuitive, since most of us have spent
a great deal of time computing Feynman diagrams. Second, it is a
better starting point towards including quantum effects. Third, it's
easier. 

\begin{myfig}{Feynman diagram giving the leading contribution to
Compton scattering of a gluon off a quark.}
\label{fig:compton}
\begin{fmfgraph*}(60,20)
\fmfleft{j,i}
\fmfright{l,k}
\fmfv{label=$k$}{i}
\fmfv{label=$p-k$}{j}
\fmf{quark}{j,x}
\fmf{quark,label=$p$}{x,y}
\fmf{quark}{y,l}
\fmf{gluon,tension=0}{i,x}
\fmf{gluon,tension=0}{y,k}
\end{fmfgraph*}
\end{myfig}

Consider Compton scattering of a gluon off a quark; see
Fig.~\ref{fig:compton}. The momentum $p$ of the internal propagator
will be  conveniently taken to have components
\[ p=(Mc+\frac1cE,\vec p).\]
This choice is the right starting point for a $1/c$ expansion; we have
written the time-like component as an energy, and shifted it by the
dominant term $Mc$. Now, we expand the intermediate propagator about
$1/c=0$; we assume the momentum $p$ ``scales'' with $c$ as given, \ie,
we take the large $c$ limit keeping $M$, $E$ and $\vec p$ fixed:
\begin{eqnarray*}
\frac1c \frac{\gamma\cdot p+Mc}{p^2-(Mc)^2} 
& = & \frac{Mc(1+\gamma_0)+\frac1cE\gamma_0-\vec\gamma\cdot\vec p}{
(Mc+\frac1cE)^2-\vec p\hskip2pt{}^2-M^2c^2}\cr\cr
&=& \frac{Mc(1+\gamma_0)+\cdots}{2ME-\vec p\hskip2pt{}^2+\cdots}\cr\cr
&=&\left(\frac{1+\gamma_0}2\right)\frac1{E-\vec p\hskip2pt{}^2/2M}+{\cal
O}(\frac1c)
\end{eqnarray*}
The factor $(1+\gamma_0)/2$ in last line projects out the $\psi$
component of the 4-spinor, and the second factor, $1/(E-\vec p\hskip2pt{}^2/2M)$,
is the non-relativistic propagator. This is precisely the propagator
for $\psi$ that follows from ${\cal L}_0$. 

For this expansion to work we need to insist that the momenta of any
quark scales with $c$ just like $p$ does, so both $p$ and $p-k$ scale
at most as $p=(Mc+\frac1cE,\vec p)$, and therefore $k$ must scale at
most as $k=(\frac1ck_0,\vec k)$.

\begin{myfig}{Quark-gluon vertex. In NRQCD the vertex is spin independent.}
\label{fig:qgq}
\begin{fmfgraph}(60,40)
\fmfleft{i}
\fmfright{k}
\fmftop{l}
\fmf{quark}{i,x,k}
\fmf{gluon,tension=0}{l,x}
\end{fmfgraph}
\end{myfig}

Consider next interactions. Since the propagators on either side of
the vertex in Fig.~\ref{fig:qgq} have a projector operator $P_+$,
where $P_\pm=(1\pm\gamma_0)/2$, $P_\pm^2=P_\pm$, $P_++P_-=1$, the
vertex can be simplified,
\[ P_+(ig_sT^a\gamma_\mu)P_+ = ig_sT^a\delta_{\mu0}\gamma_0P_+
=ig_sT^aP_+\delta_{\mu0}.
\]
Therefore only $A_0$ couples to the quark, and the coupling is a term
in the lagrangian
\[ \delta{\cal L}=\psi^\dagger g_sT^a\psi A^a_0. \]
The propagator and interaction of the quark are given therefore by the
Lagrangian
\[{\cal L}=\psi^\dagger(iD_t+\frac{\vec\nabla^2}{2M})\psi,\]
which is precisely the leading term in the quark lagrangian in the
previous section.

We can now try to construct an effective field theory in this limit. I
like to think of an effective field theory as a factorization theorem
for Green functions:~\cite{Witten}
\[
G\bigg(p_Q=(Mc+\onec E,\vec p),k_g=(\onec k^0,\vec k)\bigg)=
A(c)\tilde G\bigg((E,\vec p),(k^0,\vec k);M\bigg)+{\cal O}(\onec).
\]
There is much explaining to do. The left hand side is the Green
function for the full theory (QCD), with the kinematic variables as
chosen, \ie, in the scaling regime discussed earlier. The variables
$p_Q$ and $k_g$ are generic for the 4-momenta of quarks and gluons, of
which there may be several. The expression at right shows how the full
Green function {\it factorizes} into a term, $A(c)$, that contains all
the dependence on the small parameter, and a term that contains the
dynamical information, $\tilde G$. What makes the effective theory
useful is that $\tilde G$ can be computed as the corresponding Green
function of a theory based on an ``effective'' lagrangian.

\begin{myfig}{That a diagram like this, with an arbitrary number of
gluons, is reproduced correctly by NRQCD follows from repeated
application of the analysis for the propagator and vertex diagrams.}
\label{fig:manygluons}
\begin{fmfgraph*}(90,40)
\fmfleft{i}
\fmfright{k}
\fmftopn{l}{4}
\fmfv{label=$p$}{i}
\fmfv{label=$k_1$}{l1}
\fmfv{label=$k_2$}{l2}
\fmf{quark}{i,x1}
\fmf{quark,label=$p+k_1$}{x1,x2}
\fmf{quark,label=$p+k_1+k_2$}{x2,x3}
\fmf{quark,label=$\cdots$}{x3,x4}
\fmf{quark}{x4,k}
\fmf{gluon,tension=0}{l1,x1}
\fmf{gluon,tension=0}{l2,x2}
\fmf{gluon,tension=0,label=$\cdots$,label.dist=-140}{l3,x3}
\fmf{gluon,tension=0}{l4,x4}
\end{fmfgraph*}
\end{myfig}

In fact, we have just proved this identity at tree level for the class
of Green functions which contain one quark line and any number of
external gluons, as displayed in  Fig.~\ref{fig:manygluons}. 

Let's explore what interesting things may happen at higher order in
the loop expansion and in higher order in $1/c$, as well as for other
classes of Green functions.

\subsection{Loops}
Let us take a convergent example, like the Feynman diagram of
Fig.~\ref{fig:manygluonsloop}. This avoids unnecessary confusion from
ultraviolet divergences. We have seen that the propagators of QCD go
over into those of NRQCD if the loop momentum scales as
$k=(\frac1ck^0,\vec k)$. However this means that the gluon propagator
becomes, in leading order in $1/c$,
\begin{equation}
\label{eq:bad-glue-prop}
\frac1{\frac1{c^2}k_0^2-{\vec k}^2} \rightarrow -\frac1{\vec k^2}
\end{equation}
This is problematic, since it truncates the theory in a drastic way:
it suppresses all retardation effects. However, it seems that if we
keep the $\frac1{c^2}k_0^2$ in the propagator we fail to organize the
expansion in powers of $1/c$.

\begin{myfig}{This one loop  diagram is convergent. One must show that
NRQCD reproduces QCD's non-analytic behavior at small momenta.}
\label{fig:manygluonsloop}
\vspace{1cm}
\begin{fmfgraph*}(90,40)
\fmfleft{i}
\fmfright{k}
\fmfbottomn{l}{5}
\fmfv{label=$p$}{i}
\fmfv{label=$k_1$}{l2}
\fmfv{label=$k_2$}{l3}
\fmfv{label=$k_n$}{l4}
\fmf{quark}{i,x1,x2,x3,x4,x5,k}
\fmf{gluon,left,tension=0}{x1,x5}
\fmf{gluon,tension=0}{l2,x2}
\fmf{gluon,tension=0}{l3,x3}
\fmf{gluon,tension=0,label=$\cdots$,label.dist=70}{l4,x4}
\end{fmfgraph*}
\end{myfig}

The solution to this puzzle is to measure internal gluon momentum in
{\it energy} units. That is, we take the gluon  momentum to scale as
$k=(\frac1c k_0,\frac1c\vec k)$ where $k_0$ and  $\vec
k$ have units of energy. This is not inconsistent with our previous
scaling rules, it is just more restrictive. The propagator is then
\[\frac{c^2}{k_0^2-\vec k^2}\]
and the loop integral is over
\[\int dk_0\frac{d^3k}{c^3}. \]
This modification to our scaling rules introduces a surprise. The
internal quark propagators are now
\[\frac1{E+k_0-\frac{(\vec p +\frac1c\vec k)^2}{2M}}
=\frac1{E+k_0-\frac{{\vec p}^2}{2M}}+{\cal O}(\frac1c).
\]
In NRQCD this further expansion of the quark propagator is called a
``multipole expansion'', because it is just what we would do if the
gluon were real and we expanded in inverse powers of the large
wavelength $\lambda\sim1/|\vec k|$.  We emphasize that the previous,
tree level, success remains valid.

It is worth dwelling a bit on why we insist in retaining the $k_0$ in
the gluon propagator, rather than following our nose through the $1/c$
expansion and using the propagator in (\ref{eq:bad-glue-prop}). The
problem is that this amount to making static our fast moving modes
in the Born-Oppenheimer approximation. There are, for sure, static
components to the interaction --- in Coulomb gauge the potential $A_0$
is instantaneous. But using (\ref{eq:bad-glue-prop}) removes all
retardation effects.

\subsection{Other Green Functions}
Since we are primarily interested in bound states of heavy quark and
antiquarks, we should study almost forward $Q-\bar Q$ scattering
amplitudes; see Fig.~\ref{fig:scat-ampl}. 

\begin{myfig}{One gluon exchange. The gluon is harder than the
internal (loop) gluons in NRQCD.}
\label{fig:scat-ampl}
\begin{fmfgraph*}(60,20)
\fmfleft{j,i}
\fmfright{l,k}
\fmfv{label=$(Mc+\frac1cE,,\vec p)$}{i}
\fmfv{label=$(Mc+\frac1cE',,\vec p\hskip2pt{}')$}{k}
\fmf{plain_arrow}{i,x,k}
\fmf{plain_arrow}{l,y,j}
\fmf{curly,tension=0,
label=$\uparrow q=(\frac1c(E'-E),,\vec p\hskip2pt{}'-\vec p)$,
label.dist=-1000}{x,y}
\end{fmfgraph*}
\end{myfig}

The external initial and final quark momenta are chosen according to
our scaling rules, $p=(Mc+\frac1cE,\vec p)$ and $p'=(Mc+\frac1cE',\vec
p\,{}')$.  This implies the momentum transfer
$q=p\,{}'-p=(\frac1c(E'-E),{\vec p}\,{}'-\vec p)$, carried by the gluon
has a spatial component that scales as $c^0$ rather than $1/c$. Is this a
problem?

\begin{myfig}{Local interaction reproducing the hard gluon exchange 
of Fig.~\ref{fig:scat-ampl}.}
\begin{fmfgraph*}(60,20)
\fmfleft{j,i}
\fmfright{l,k}
\fmf{quark}{i,x,k}
\fmf{quark}{l,x,j}
\fmfv{decoration.shape=circle,decoration.size=50,decoration.filled=1}{x}
\end{fmfgraph*}
\end{myfig}

Our rules state that the internal integration variables for gluon
momenta should have units of energy, and therefore come accompanied by
appropriate factors of $1/c$. But in the diagram of
Fig.~\ref{fig:scat-ampl} the momenta in the gluon propagator is fixed
externally and scales as $c^0$, while the energy scales as $1/c$. 
So the amplitude is

\[ 
\parbox{20mm}{
\begin{fmfgraph*}(60,20)
\fmfleft{j,i}
\fmfright{l,k}
\fmf{quark}{i,x,k}
\fmf{quark}{l,x,j}
\fmfv{decoration.shape=circle,decoration.size=50,decoration.filled=1}{x}
\end{fmfgraph*}}\qquad\qquad\qquad\qquad\qquad\qquad  
\sim\qquad-\frac1{{\vec q}\hskip2pt^2} \cdot g_s^2 T^a\otimes T^a~.
\]
This can be reproduced by our effective theory as a spatially
non-local 4-quark vertex. Since the Fourier transform of $1/|\vec q|^2$
is $1/|\vec x|$, the effective lagrangian should be augmented by terms
that are schematically of the form
\[ \psi^\dagger\psi(t,\vec x)\int d^3y\frac1{|\vec x-\vec y|}
\chi^\dagger\chi(t, \vec y).\]
This interaction is responsible for, among others, the Coulomb-like
potential between $Q$ and $\bar Q$.

Note that the interaction is local in time, as it must for an
appropriate Hamiltonian formulation of the problem.

Exercise: Compute explicitly the Coulomb term that needs to be added
to the effective theory.

\subsection{The NRQCD Lagrangian Revisited}
After our exploration in momentum space we return to a derivation of
the effective lagrangian directly in configuration space.  Our aim is
to understand in terms of a Lagrangian formulation the origin of the
new rules that we derived by considering Feynamn diagrams.  From here
on we will adopt Coulomb gauge $\vec \nabla\cdot \vec A=0$. As we saw
earlier the lowest order lagrangian density
\[
{\cal L}_0=
\frac12(\partial_iA^a_0-\frac1c\partial_tA_i^a)^2
-\frac12(\partial_iA^a_j)^2
+\psi^\dagger(iD_t+\frac{\vec \nabla^2}{2M})\psi 
\]
still contains $c$ dependence. Defining~\cite{GR}
\[ \tilde A_i(\vec y =\frac{\vec x}c,t)=\sqrt{c}A_i(\vec x,t)
\]
and writing the lagrangian in terms of this new variable, we obtain
\[ L_0=\int
d^3y\left[\frac12(\partial_0\tilde A_i)^2-\frac12(\partial_i\tilde
A_j)^2\right]
+\int d^3x\left[\psi^\dagger(iD_t+\frac{\vec\nabla^2}{2M})\psi
+\frac12(\partial_iA_0)^2\right]\]
This is independent of $c$ and can be used as a starting point for an
expansion in $1/c$. 

In particular:
\begin{itemize}
\item The states of $L_0$ are $c$-independent
\item These states can be used to formulate perturbation theory in
$1/c$
\item All $1/c$ corrections can be expressed as matrix elements of
operators between these states. That is, we have {\it explicit power
counting}. The engineering dimensions of an operator can be
determined, and compensated for with the appropriate inverse powers of
mass, $M$, and speed of light, $c$.
\end{itemize}

Consider the $1/c$ corrections. The terms of order $1/c$ in the
Lagrangian density of Eq.~(\ref{eq:lag-NRQCD}) are
\[
L_{c^{-1}}=\int d^3x\;\psi^\dagger(\vec x,t)
\left[\frac{g}{Mc^{3/2}}\tilde A_i(\onec\vec
x,t)\frac\partial{\partial x^i}
+\frac{c_Fg}{2Mc^{5/2}}\vec\sigma\cdot\tilde {\bf B} (\onec\vec x,t)
\right] \psi(\vec x,t),
\]
where \(\tilde B_i =\epsilon_{ijk}\frac\partial{\partial y^j}
\tilde A_k(\vec y,t)\). As we see, in these interaction terms the
fields have dependence both on $\vec x$ and on $\vec y=\frac1c\vec
x$. Changing variables to $\vec y$ does not eliminate the dependence on
$c$. Instead we write all field dependence  in terms of $\vec x$ and
expand in a Taylor  series the terms that depend on $\frac1c\vec
x$. This is just the multipole expansion:~\cite{GR,Labelle}
\begin{eqnarray}
L_{c^{-1}{\rm MP}}=\int d^3x\;\psi^\dagger(\vec x,t)& & 
\left\{\frac{g}{Mc^{3/2}}\left[\tilde A_i(\vec 0,t) +
\frac{x_j}c(\partial_j\tilde A_i(\vec 0,t) +\cdots)\right]
\frac\partial{\partial x^i}\right. \nonumber \\
& &\left.+\frac{c_Fg}{2Mc^{5/2}}\left[\vec\sigma\cdot\tilde {\bf B}(\vec
0,t)+\cdots\right]\right\} \psi(\vec x,t). \label{eq:MPlag}
\end{eqnarray}
Although this expression seems unfamiliar, it is the configuration
space version of the multipole expansion performed earlier in momentum
space were it appeared in a natural and intuitive form.

\subsection{Technical Note On Explicit Power Counting}
The multipole expansion makes power counting straightforward, and this
is why we have introduced it here. Let us see how it works in more
detail in a somewhat non-trivial example.

\begin{myfig}{Contribution to the quark self-energy from two insertions
of the operator $\psi^\dagger\vec\sigma\cdot\vec B\psi$, represented
by a shaded circle. Without a multipole expansion this graph gives
contributions of lower order in $1/c$ than the order of the operators
inserted.}
\begin{fmfgraph}(80,40)
\fmfleft{i}
\fmfright{f}
\fmf{quark}{i,x,y,f}
\fmf{gluon,left,tension=0}{x,y}
\fmfblob{50}{x}
\fmfblob{50}{y}
\end{fmfgraph}
\end{myfig}

Consider a contribution to the quark self-energy from two insertions
of the operator $\psi^\dagger\vec\sigma\cdot\vec B\psi$. This is one
of the terms of order $1/c$ in the lagrangian density in
Eq.~(\ref{eq:lag-NRQCD}). 

If we use this operator in the lagrangian as it stands in
Eq.~(\ref{eq:lag-NRQCD}), that is, before multipole expanding, the one
loop contribution to the dimensionally regularized two-point function is
\[
i\Gamma^{(2)} =
\frac{c_F^2C(R)g^2}{4M^2c^2}\int\frac{d^Dk}{(2\pi)^D}
\frac{(\vec k\times\vec\sigma)\cdot(\vec k\times\vec\sigma)}%
{\left(\frac1{c^2}k_0^2-\vec k^2+i\epsilon\right)
\left(E+k_0-\frac{(\vec p + \vec k)^2}{2M}+i\epsilon\right)}.
\]
With $D=4-2\varepsilon$, the divergent pieces as $\varepsilon\to0$ are, 
\[
i\Gamma^{(2)}_{{\rm div}} = \frac{c_F^2C(R)\alpha_s}\pi \Gamma(2\varepsilon)
\left[2\left(E-\frac{\vec p\hskip2pt^2}{2M}\right)+
4Mc^2 +\frac43\frac{\vec p\hskip2pt^2}{2M}\right].
\]
Other schemes yield analogous results.

As an exercise the reader should compute the loop integral with a
cut-off\break $|\vec k|<\Lambda$. Clearly he or she will find a term of order
\(\frac{c_F^2C(R)\alpha_s}\pi\Lambda^2\). This result, and others like
it, has led many (including me) to make the ridiculous statement
that the effective theory must be defined with a {\it low} cut-off,
say, \(\Lambda\sim{\rm Rydberg}\sim1/Ma^2\).

This is, of course, non-sense. One should take the cutoff $\Lambda$ to
be arbitrarily large, and make the necessary subtractions of divergent
and finite terms. However making subtractions is a bit of a
pain. Firstly, since the lower order terms in the $1/c$ expansion are
modified by the inclusion of any higher order term, one must
recalculate counterterms every time a new higher order term is
added. Secondly, this means that to restore power counting one must
tune coefficients carefully.

If instead one computes with the multipole expanded lagrangian, the
result in dimensional regularization is
\[
i\Gamma^{(2)} =
\frac{c_F^2C(R)g^2}{2M^2c^5}\int\frac{d^Dk}{(2\pi)^D}
\frac{\vec k^2}%
{\left(k_0^2-\vec k^2+i\epsilon\right)
\left(E-\frac{\vec p^2}{2M}+k_0+i\epsilon\right)}.
\]
Note that, by design, the whole dependence on $c$ is explicit in the
coefficient, $1/c^5$. The divergent part is now
\[
i\Gamma^{(2)}_{{\rm div}} = \frac{c_F^2C(R)\alpha_s}{2\pi}
\Gamma(2\varepsilon) \frac{(E-\vec p\hskip2pt^2/2M)^3}{M^2c^4}.
\]

\subsection{NRQCD -- An Effective Field Theory: Is All This Really Necesary?}
Although nobody has bothered to proof the validity of NRQCD as an
effective theory (in the sense of a factorization theorem, as
explained above), we believe this to be true. Why?

The answer is that NRQCD has the right ingredients. That is, one could
simply ask what terms should be contained in  a local lagrangian field
theory describing gluons and non-relativistic quarks.  The
answer~\cite{LMNMH} is the lagrangian we have been considering,
\begin{equation}
\label{eq:efflag}
{\cal L}=
-\frac14F^a_{\mu\nu}F^{a\mu\nu}
+\psi^\dagger(iD_t+\frac1{2M}\vec D^2)\psi + 
\frac{c_Fg}{2Mc}\psi^\dagger\vec\sigma\cdot\vec B\psi +\cdots
\end{equation}
The coefficient of the $F^2$ and $\psi^\dagger D_t\psi$ terms have
been scaled to one by wave function renormalization, as usual. The
coefficient of the $\psi^\dagger \vec D^2\psi$ term {\it defines} the
mass $M$. Coefficients of higher dimension operators
($c_F,c_D,\ldots$) remain undetermined. In a non-perturbative
treatment of the theory, as for example on the lattice, one can treat
$c_F, c_D, c_S$ as additional free couplings to be determined by
comparing with experiment (or with full, continuum QCD, in which case
we call the procedure ``matching'').

It would appear then that all the careful analysis of the previous
sections, and in particular the multipole expansion, is really
unnecessary. The lagrangian of Eq.~(\ref{eq:efflag}) describes the same
physics as our multipole expanded lagrangian of
Eq.~(\ref{eq:MPlag}). The only difference is that in the first case
one may not systematically study each order of the $1/c$
expansion. This may not be important for many applications, and in
such cases the simpler formalism should be employed. But, as we will
see, this is not always the case. An important example is given in the
next chapter. The calculation of the decay of $P$-wave quarkonium
exhibits infrared divergences in the matching to the effective
theory that can be eliminated properly only if one includes the
correct set of operators. Counting of powers of $1/c$ singles out the
correct operators. Moreover, if some of the $1/c$ effects were left
implicit in the lowest order states (or elsewhere) one could in
principle have some of the infrared cancellation take place against
part of the state definition: this would make such a cancellation
practically intractable.

\subsection{Spin Symmetry}
\label{sec:spinsymmetry}
The Lagrangian
\[
{\cal L}=\psi^\dagger(iD_t+\frac1{2M}\vec D^2)\psi 
-\chi^\dagger(iD_t+\frac1{2M}\vec D^2)\chi 
\]
has a $U(2)\times U(2)$ symmetry
\[
\psi_\alpha\to R_{\alpha\beta}\psi_\beta\qquad\qquad
\chi_{\dot\alpha}\to T_{\dot\alpha\dot\beta}\chi_{\dot\beta}
\]
with \(R^\dagger R=T^\dagger T={\bf 1}\). Please note that while the
field $\psi$ annihilates a heavy quark, the field $\chi$ creates a
heavy anti-quark. The abelian factors $U(1)\times U(1)$ correspond to
separate conservation of heavy quark and anti-quark numbers. The
$SU(2)$'s are {\it spin symmetries}. They are broken by the spin-flip
magnetic moment interaction that appears in $L_{c^{-1}}$, and which is of
order $c^{-5/2}$ in the multipole expansion:
\[
L_{c^{-1}{\rm MP}}=\int d^3x\;\frac{c_Fg}{2Mc^{5/2}}\psi^\dagger(\vec x,t)
\vec\sigma\cdot\tilde {\bf B}(\vec
0,t) \psi(\vec x,t)+\cdots
\]
None of these observations  come as a surprise to the reader who has
kept in mind the molecular physics analogue of quarkonium.

In non-relativistic quantum mechanics orbital angular momentum $\vec
L$ and spin $\vec S$ are separately conserved, and therefore so is the
total angular momentum $\vec J=\vec L+\vec S$. States are specified by
${}^SL_J$, where in the standard spectroscopic notation $S$ takes
values denoting the dimension of the $SU(2)$ representation ($S=1$ for
the singlet, $S=3$ for triplet (adjoint), etc), $L= S, P, D, \ldots$ and
$J=0,1,2,\ldots$ In the following table we give the possible
quark--anti-quark states for the lowest few values of $L$, and the
name of the state for the case of charm quarks:
\begin{displaymath}
\begin{array}{|lcc|}\hline
L=0 & & \\
  & {}^1S_0 & \eta_c \\
  & {}^3S_1 & J/\psi \\ \hline
L=1 & & \\
 & {}^1P_1& h_c \\
 & {}^3P_0&\chi_{c0}\\
 & {}^3P_1 &\chi_{c1}\\
 & {}^3P_2&\chi_{c2}\\ \hline
\end{array}
\end{displaymath}
In leading order in the $1/c$ expansion states are classified this
way, and their properties are related by spin symmetry. The names of
b-quark states are obtained by replacing $c\to b$, except for the
${}^3S_1$ which is called  ``$\Upsilon$.''

The spin symmetry is most easily applied by use of the Wigner-Eckart
theorem. This is best explained through examples.

\subsubsection{Example 1: S-states}
Construct $2\times2$ matrix $H_{\alpha\dot\beta}$ representing  the four
\({}^1S_0, {}^3S_1\) states. Under spin symmetry
\[ H\rightarrow RHT^\dagger.
\]
Then, for any matrix $\Gamma$ and any state $X$ free of heavy quarks
(and therefore a singlet under spin symmetry), 
the Wigner-Eckart theorem gives
\[
\langle X|\chi^\dagger\Gamma\psi|H\rangle =\xi_{\scriptscriptstyle X}{\rm Tr}\Gamma H.
\]
Here $\xi_X$ is the reduced matrix element; generally it depends on
kinematic variables and the parameters of the theory. We have used the
same letter $H$ to denote the specific quarkonium state on the left
hand side and the corresponding matrix on the right. Explicitly (for
$b$-quarkonium) 
\[
H=\sigma^0\eta_b+\sum_{a=1}^3\sigma^a\Upsilon^a.
\]
Here and below $\sigma^0={\rm diag}(1,1)$.

This implies, for example
\begin{eqnarray*}
\langle X|\chi^\dagger\psi|\eta_b\rangle & =&\xi_{\scriptscriptstyle X}\\
\langle X|\chi^\dagger\sigma^a\psi|\Upsilon(\epsilon)\rangle 
& =&\xi_{\scriptscriptstyle X}\epsilon^a\\
\end{eqnarray*}
where $\vec\epsilon$ is the polarization vector of the spin-1
$\Upsilon$. 
\subsubsection{Example 2: P-states}
More interesting is the case of the 12 P-wave states. The
$12=3\times2\times2$ states can be represented by 
\[\tilde H_{m\alpha\dot\beta}, \qquad m=1,2,3\qquad\alpha,\dot\beta =
1,2
\]
and the components are found by insisting they transform correctly
under rotations (index $m$) and spin symmetry ($\alpha$ and
$\dot\beta$), combined into appropriate $J$. Without proper normalization:
\[
\tilde H_{m\alpha\dot\beta} = \sigma^m_{\alpha\dot\beta}\chi_0
+\epsilon_{mjk}\sigma^j_{\alpha\dot\beta}\chi^k_1
+\sigma^j_{\alpha\dot\beta}\chi_2^{mj}
+\sigma^0_{\alpha\dot\beta}h_m
\]
Here \(\chi_2^{mj}\) is  symmetric and traceless.

Exercise: (a) Normalize these properly. To this end compute ${\rm Tr}
H^\dagger H$ and verify that the coefficients of all states are
unity.\newline
(b)In the rest frame of the quarkonium state ($\vec p=0$) what is
\(\langle0| \chi^\dagger\Gamma\psi|\tilde H\rangle\)? Calculate 
\(\langle0| \chi^\dagger
\darr D_m\Gamma
\psi|\tilde H\rangle\). Find the relations between these matrix
elements of $\chi_J$ and $h$.

\section{Annihilation Decays of Quarkonium}
\subsection{Color Singlet Model}
Consider the decay of $\eta_c$ into light hadrons. To this effect
model the state by a pure singlet,
$\eta_c=(Q\bar Q)_1$, that is, with no $(Q\bar Q)_8$ component. Let us
represent the
$Q\bar Q$ in the ${}^1S_0$ state as in the quark model
\[
|\eta_c\rangle=\frac1{\sqrt{2M_\eta}}\int\frac{d^3q}{(2\pi)^3}
\psi(\vec q)\frac{\delta^{ij}}{\sqrt3}\frac{\epsilon^{mn}}{\sqrt2}
|c^{im}(\vec q)\bar c^{jn}(-\vec q)\rangle,
\]
where $i,j$ are color indices, $m,n$ are spin indices and $\psi(\vec
q)$ is the Fourier transform of the coordinate space wavefunction
\[
\psi(\vec x) = \int\frac{d^3q}{(2\pi)^3} e^{i\vec q\cdot\vec x} 
\psi(\vec q).
\]
By spherical symmetry
\[
\psi(\vec x) = \frac1{\sqrt{4\pi}}R(r)
\]
where $r=|\vec x|$.

We stop here for a small digression. We note that something is
missing. Even in the molecular physics analogy of  Sect.~1 the state
would be represented by a wave-function $\phi_n(x,X)\psi(X)$, with
$\phi_n(x,X)$ the wavefunction for the fast degrees of freedom. We are
missing the wavefunction for the glue! We will look the other way and
carry on, as if the glue component were trivial.

\begin{myfig}{Feynman diagrams for the  $cc\to gg$
amplitude, used in the calculation of the $\eta_c$ width.}
\label{fig:cctogg}
\[\parbox{20mm}{
\begin{fmfgraph*}(60,20)
\fmfleft{j,i}
\fmfright{l,k}
\fmfv{label=$q$}{i}
\fmfv{label=$q'$}{j}
\fmfv{label=$k,,A$}{k}
\fmfv{label=$B$}{l}
\fmf{quark}{i,x}
\fmf{quark,tension=0,label=$q-k$}{x,y}
\fmf{quark}{y,j}
\fmf{gluon}{x,k}
\fmf{gluon}{y,l}
\end{fmfgraph*}}\qquad\qquad\qquad\qquad\qquad\qquad
+\hbox{crossed}+\hbox{higer order}
\]
\end{myfig}

In order to calculate the total inclusive width
$\Gamma(\eta_c\to{\rm hads})$ we use {\it duality}. This is the
physically plausible assumption that the  sum  over all  final hadronic
states should be well approximated by the rate into partons,
\[
\Gamma(\eta_c\to{\rm hads})=\sum_X \Gamma(\eta_c\to X)\approx
\Gamma(\eta_c\to gg).
\]
With our representation of the $\eta_c$ state we then have
\[
\Gamma(\eta_c\to{\rm hads})
=\frac1{2M_\eta}\int\frac{d^3k}{(2\pi)^32k_0}
\frac{2\pi\delta(M_\eta-2|\vec k|)}{M_\eta}
|{\cal M}(\eta_c\to g(\vec k)g(-\vec k)|^2.
\]
Here ${\cal M}$ is the invariant (T-matrix) amplitude. Suppressing
color and spin indices
\[
{\cal M}(\eta_c\to g(\vec k)g(-\vec k)) =
\frac1{\sqrt{2M_\eta}}\int\frac{d^3q}{(2\pi)^3}\psi(\vec q)
{\cal M}(c(\vec q)\bar c(-\vec q)\to g(\vec k)g(-\vec k)).
\]
In perturbation theory the amplitude ${\cal M}(c(\vec q)\bar c(-\vec
q)\to g(\vec k)g(-\vec k))$ is computed from the diagram in
Fig.~\ref{fig:cctogg}, and is given by
\[
{\cal M}(c(\vec q\,)\bar c(-\vec
q\,)\to g(\vec k)g(-\vec k)) =
-ig_s^2(T^BT^A)
\bar v\gamma\cdot\epsilon^*(k')
\frac1{\gamma\cdot(q-k)-m_c}
\gamma\cdot\epsilon^*(k)u
\]
where $q=(m_c+E,\vec q)$, $q'=(m_c+E,-\vec q)$ and $k=\frac12M_\eta(1,\hat n)$
($|\hat n|=1$). Note that, for now, I have suppressed powers of $c$
which can be restored by dimensional analysis. The denominator
\begin{equation}
\label{eq:kinemats}
(q-k)^2-m_c^2\approx-2q\cdot k=-M_\eta[(m_c+E)-\vec q\cdot\hat n]
\end{equation}
is dominated by the mass term for $|\vec q|<m_c$ and there it is
approximately constant (or
``flat'') and $\approx-2m_c^2$. The wavefunction $\psi(\vec q)$ has
support within one inverse Bohr radius, $1/a_{\rm
Bohr}\sim\alpha_sm_c$, which is indeed smaller than $m_c$ if a
perturbative value for $\alpha_s$ can be used.

\begin{myfig}{The amplitude for $cc\to gg$ is effectively local: when
the external charm quarks are non-relativistic the internal quark must
be off shell by an amount of order of the charm mass.}
\label{fig:cctoggeff}
\[
\hskip-50mm\parbox{20mm}{
\begin{fmfgraph}(60,20)
\fmfleft{j,i}
\fmfright{l,k}
\fmf{quark}{i,x}
\fmf{quark,tension=0}{x,y}
\fmf{quark}{y,j}
\fmf{gluon}{x,k}
\fmf{gluon}{y,l}
\end{fmfgraph}
}
\qquad\qquad\qquad\qquad\qquad
\longrightarrow
\parbox{20mm}{
\begin{fmfgraph}(60,20)
\fmfleft{j,i}
\fmfright{l,k}
\fmf{quark}{i,x}
\fmf{quark}{x,j}
\fmf{gluon}{x,k}
\fmf{gluon}{x,l}
\fmfv{decoration.shape=circle,decoration.size=50,decoration.filled=1}{x}
\end{fmfgraph}
}
\]
\end{myfig}

We therefore approximate the integral by replacing the propagator by
its flat value,
\begin{eqnarray*}
\Gamma(\eta_c\to{\rm hads})& \approx&
\frac1{(2M_\eta)^2}\left|\int\frac{d^3q}{(2\pi)^3}\psi(\vec
q)\right|^2\cr
&\times&
\int\frac{d^3k}{(2\pi)^32k_0}
\frac{2\pi\delta(M_\eta-2|\vec k|)}{M_\eta}
|{\cal M}(c(0)\bar c(0)\to g(\vec k)g(-\vec k)|^2\cr
\end{eqnarray*}

Exercise: Complete the calculation. Show this gives
\[
\Gamma(\eta_c\to{\rm hads})\approx\frac{8\alpha_s^2}{3M_\eta^2}
|R(0)|^2.
\]

\subsection{$\eta_c$ Decay in NRQCD}
We would like to carry out the calculation of the previous section in
a more rigorous manner. To this end we would like to use the effective
theory, NRQCD. However, several problems immediately come to mind:
\begin{itemize}
\item The effective theory cannot describe
properly gluons or quarks with $E,|\vec p|\sim m_c$ (because the $1/c$
expansion will result in an expansion in velocity $v$ with $v/c\sim
|\vec p|/m_cc$ or $v^2\sim E/m_cc^2)$.
\item The effective theory has separate conservation of $Q$ and $\bar
Q$ numbers. How can it possibly describe $Q\bar Q$ annihilation in
$\eta_c$ decay?
\end{itemize}

\begin{myfig}{Feynman diagram for the forward $c\bar c$ scattering amplitude.}
\label{fig:ccbox}
\begin{fmfgraph*}(60,20)
\fmfleft{j,i}
\fmfright{l,k}
\fmfv{label=$p$}{i}
\fmfv{label=$p\hskip2pt{}'$}{j}
\fmfv{label=$p$}{k}
\fmfv{label=$p\hskip2pt{}'$}{l}
\fmf{quark}{i,x}
\fmf{quark,tension=0,label=$p+k$}{x,y}
\fmf{quark}{y,j}
\fmf{quark}{w,k}
\fmf{quark,tension=0,label=$k-p$}{z,w}
\fmf{quark}{l,z}
\fmf{gluon,label=$k$}{x,w}
\fmf{gluon,label=$k+p-p\hskip2pt{}'$}{y,z}
\end{fmfgraph*} 
\end{myfig}

Fortunately there is a common solution to these
problems.\cite{BBYL,BBL} Consider the Feynman diagram for $c\bar c$
annihilation in Fig.~\ref{fig:cctogg}. The external quark lines,
representing the quark-antiquark pair in charmonium, are almost
on-shell. But the internal quark line is off the mass shell by the
mass of the quark itself! Effectively the interaction is local; see
Fig.~\ref{fig:cctoggeff}. Formally the denominator in
Eq.~(\ref{eq:kinemats}) can be expanded
\[
\frac1{m_c+E-\vec q\cdot\hat n}=\frac1{m_c}\left(1 -
                 \frac{E-\vec q\cdot\hat n}{m_c}+\cdots\right)
\]
and the leading term gives a local non-derivative quark-gluon
interaction for $c\bar c\to gg$. 

\begin{myfig}{The optical theorem relates the total $c\bar c$
annihilation cross section, which enters the calculation of the width
of the $\eta_c$, to the $c\bar c$ forward scattering amplitude, which
can be modeled by a local interaction in NRQCD.}
\[
\hskip-0.4cm\left|
\parbox{20mm}{
\begin{fmfgraph}(60,20)
\fmfleft{j,i}
\fmfright{l,k}
\fmf{quark}{i,x}
\fmf{quark,tension=0}{x,y}
\fmf{quark}{y,j}
\fmf{gluon}{x,k}
\fmf{gluon}{y,l}
\end{fmfgraph}}
\hskip4cm\right|^2
=2\hbox{Im}
\left(
\parbox{20mm}{
\begin{fmfgraph}(60,20)
\fmfleft{j,i}
\fmfright{l,k}
\fmf{quark}{i,x}
\fmf{quark,tension=0}{x,y}
\fmf{quark}{y,j}
\fmf{quark}{k,w}
\fmf{quark,tension=0}{w,z}
\fmf{quark}{z,l}
\fmf{gluon}{x,w}
\fmf{gluon}{y,z}
\end{fmfgraph}}
\hskip4cm\right)
\]
\end{myfig}

Notice however that the external gluons have energy and momenta of
order $m_c$. This, of course, is not appropriate for treatment in the
effective theory. So we are not out of the woods yet. The proposed solution
is to use the optical theorem which gives the rate for $c\bar c\to gg$
from the imaginary part of the forward $c\bar c$ scattering amplitude.
The lowest order Feynman diagram for $c\bar c\to c\bar c$ is shown is
Fig.~\ref{fig:ccbox}. The imaginary part has the internal lines off
the mass shell by about $m_c$, so the whole loop is local on the scale
of $|\vec q|\sim a_{{\rm Bohr}}^{-1}\sim m_c\alpha_s\sim m_cv$ and we
may replace it by a local interaction. This is shown in
Fig.~\ref{fig:ccboxeff} where $\kappa$ is a constant and the ellipsis
indicate terms of order $|\vec q|/m_c$. The constant $\kappa$ can be
easily computed in the limit $\vec q=0$.


\begin{myfig}{The $c\bar c$ forward scattering amplitude
can be modeled by a local interaction in NRQCD. The constant $\kappa$
is determined by ``matching'' to QCD.}
\label{fig:ccboxeff}
\[
\hskip-0.3cm
\parbox{20mm}{
\begin{fmfgraph}(60,20)
\fmfleft{j,i}
\fmfright{l,k}
\fmf{quark}{i,x}
\fmf{quark,tension=0}{x,y}
\fmf{quark}{y,j}
\fmf{quark}{w,k}
\fmf{quark,tension=0}{z,w}
\fmf{quark}{l,z}
\fmf{gluon}{x,w}
\fmf{gluon}{y,z}
\end{fmfgraph}}
\hskip4cm\approx
\kappa
\parbox{20mm}{
\begin{fmfgraph}(60,20)
\fmfleft{j,i}
\fmfright{l,k}
\fmf{quark}{i,x,j}
\fmf{quark}{l,x,k}
\fmfv{decoration.shape=circle,decoration.size=50,decoration.filled=1}{x}
\end{fmfgraph}}\hskip3.3cm+\cdots
\]
\end{myfig}

To get some better understanding of how this works, let's look at the
propagators in the box diagram of Fig.~\ref{fig:ccbox}. Let $k$
stand for the loop momentum and $p$ and $p\,{}'$ the external momenta. The
quark propagators give
\[ 
\frac1{k^2+2p\cdot k}\frac1{k^2-2p\cdot k}
\]
and the gluon propagators 
\[
\frac1{k^2}\frac1{(k+p-p\,{}')^2}.
\]
Now, the imaginary part of the diagram gives a cut on the gluon lines,
\[
{\rm Im}\frac1{k^2+i\epsilon}=-\pi\delta(k^2)
\]
so the product of internal propagators reduces to
\[
-\frac{\pi^2}{4(p\cdot k)^2}\delta(k^2)\delta((k+p-p\,{}')^2).
\]
This is easy to interpret: the two gluons are on shell, the two quark
propagators are just as before, Eq.~(\ref{eq:kinemats}).

In NRQCD the annihilation process is modeled by adding to the
lagrangian a non-hermitian term that gives the imaginary part of the
$Q\bar Q$ forward scattering amplitude. That is, the lagrangian is
augmented by a local four-fermion operator
\[
\Delta{\cal L}=\frac{f}{M^2c}\psi^\dagger\chi\chi^\dagger\psi
\]
with $f$ a dimensionless\hskip2pt\footnote{To count dimensions, recall
$S=\int dt\,d^3x {\cal L}$ is dimensionless, $ {\cal L} =
\psi^\dagger\partial/\partial t\; \psi$, so $[\psi^\dagger\psi]\sim
L^{-3}$. Since $\hbar=1$, \([Mc]\sim L^{-1}\) so we have \( [\int
dt\,d^3x \frac{c}{(Mc)^2}\psi^\dagger\chi\chi^\dagger\psi] \sim L^4
L^2 L^{-6}\sim1\).} constant chosen judiciously to give the right
answer to this order in the $1/c$ expansion. This operator is the
product of two $(^1S_0)_1$ bilinears (the subindex refers to the color
channel, the spin quantum numbers are inside the parenthesis). The
first bilinear annihilates the $(^1S_0)_1$ $c\bar c$, while the second
creates it from the vacuum.

The constant $f$ is determined by ``perturbative matching''. This
means that the rate should correspond to that of full QCD to the
accuracy of the order of relevance of the NRQCD approximation. But the
same constant $f$ should reproduce the perturbative QCD calculation of
a partonic cross section. Let's consider this procedure in some
detail. In QCD the forward scattering amplitude is given by the box
diagram of Fig.~\ref{fig:ccbox}. The spinors come in as
\[
\bar v \gamma^\mu[\gamma\cdot(p+k)+m]\gamma^\nu u\,
\bar u \gamma^\nu[\gamma\cdot(p-k)+m]\gamma^\mu v.
\]
In the NR-limit we can replace
\[ u=\pmatrix{\psi\cr0\cr}\qquad\hbox{and}\qquad\bar v =
\pmatrix{0\cr-\chi^\dagger\cr}
\]
The product of  Dirac matrices is reduced by 
$\gamma^\mu\gamma^\lambda\gamma^\nu=\gamma^\mu\eta^{\lambda\nu}
+\gamma^\nu\eta^{\lambda\mu}-\gamma^\lambda\eta^{\mu\nu}-
i\epsilon^{\mu\nu\lambda\rho}\gamma_\rho\gamma_5$
and $\gamma^\mu\gamma^\nu=\eta^{\mu\nu}-i\sigma^{\mu\nu}$. The
needed binomials are
\begin{eqnarray*}
\bar v \gamma^\mu u &=& \cases{0&$\mu=0$\cr\chi^\dagger\sigma^i\psi&$\mu=i$}\\
\bar v \gamma^\mu \gamma_5u &=& 
\cases{\chi^\dagger\psi&$\mu=0$\cr0&$\mu=i$}\\
\bar v \sigma^{0i} u  &=& -i\chi^\dagger\sigma^i\psi\\
\bar v \sigma^{ij} u  &=& 0
\end{eqnarray*}
Notice that when we construct singlet combinations we have, in
addition,
\[
\sum_{{\rm singlet}} \chi^\dagger\sigma^i\psi =0.
\]
Therefore the only non-vanishing singlet-singlet contribution is from
\[
 i\epsilon^{\mu\lambda\nu0}i\epsilon_{\mu\sigma\nu0}
(p+k)_\lambda (p-k)^\sigma \chi^\dagger\psi\psi^\dagger\chi
\]
The coefficient is just \(
 [(p+k)\cdot(p-k)-(p+k)^0(p-k)^0] \)
and this, on-shell (when we  take the imaginary
part of the amplitude), is \(m_c^2 \).

Exercise: Complete the calculation. Show that the spin singlet, color
singlet part of the cut diagram is
\[
\frac{2\pi\alpha_s^2}{9m_c^2}\chi^\dagger\psi\psi^\dagger\chi.
\]

In NRQCD this is to be reproduced by a term
\[
\frac{f}{m_c^2}\chi^\dagger\psi\psi^\dagger\chi.
\]
in the Lagrangian. The symbols \(\chi\) and \(\psi\) are fields, but
the tree level amplitude corresponds to replacing them by
spinors. The matching condition is
\[
{\rm Im}\, f = \frac{2\pi\alpha_s^2}{9}
\]

Finally, we compute the rate $\Gamma(\eta_c)$ in NRQCD. Treating
${\cal L}$ as a perturbation, the energy shift is\hskip2pt\footnote{In
the standard formula \(\Delta
E=\frac{\langle\psi|H'|\psi\rangle}{\langle\psi|\psi\rangle}\) go to
relativistic normalization:
\(\langle\psi|\psi\rangle=2M_{\eta_c}\delta^3(0)=2M_{\eta_c}V\). The
factor of $V^{-1}$ is used in $H'V^{-1}={\cal H}'=-\Delta{\cal L}$.}
\[
\Delta E_{\eta_c} =-\frac{f}{m_c^2c}
\frac{\langle\eta_c| \psi^\dagger\chi\chi^\dagger\psi |\eta_c\rangle}%
{2M_{\eta_c}}.
\]
The imaginary part of $\Delta E$ is just $\frac12\Gamma$, so we have
finally
\[
\Gamma(\eta_c)=\frac1{2M_{\eta_c}}\frac{4\pi\alpha_s^2}{9m_c^2}
\langle\eta_c| \psi^\dagger\chi\chi^\dagger\psi | \eta_c\rangle.
\]

We can find the relation between this result and that of the
color-singlet model by using the vacuum insertion approximation
\[
\langle\eta_c| \psi^\dagger\chi\chi^\dagger\psi | \eta_c\rangle
\approx 
|\langle0| \chi^\dagger\psi | \eta_c\rangle|^2
\]
with
\[
\langle0| \chi^\dagger\psi | \eta_c\rangle\approx\sqrt{2M_{\eta_c}}\;
\sqrt\frac3{2\pi}\;R(0)
\]

We close this section with a short digression. The reader may be
concerned with the appearance of complex coefficients of hermitian
operators in the lagrangian. This renders the Hamiltonian
non-hermitian and jeopardizes conservation of probability. A little
thought shows that this is as expected. The situation is similar to
that encountered in the effective two level Hamiltonian for $K^0-\bar
K^0$ mixing. There the Hamiltonian is given in terms of hermitian
matrices $M$ and $\Gamma$ as $H=M+\frac i2\Gamma$. The loss of
hermiticity encoded in $\Gamma$ arises because the states into which
the $K$-mesons may decay are not incorporated in the
theory. Similarly, in NRQCD we have truncated the theory by excluding
energetic gluons and light quarks, but these are the decay products of
heavy quarkonia.

\begin{myfig}{Feynman diagram for $c\bar c$ annihilation entering the
calculation of the $\eta_c$ decay width.}
\label{fig:ccggagain}
\begin{fmfgraph}(60,20)
\fmfleft{j,i}
\fmfright{l,k}
\fmf{quark}{i,x}
\fmf{quark,tension=0}{x,y}
\fmf{quark}{y,j}
\fmf{gluon}{x,k}
\fmf{gluon}{y,l}
\end{fmfgraph} 
\end{myfig}

\eject
\subsection{Radiative Corrections and Decays of P-Waveonium}
In the $c\bar c$ annihilation diagram of Fig.~\ref{fig:ccggagain},
used to describe $\eta_c$ decay, the gluons have momentum $|\vec
k|=\frac12M_{\eta_c}$. Consider now the radiative corrections shown in
the diagrams of Fig.~\ref{fig:softg}. We
are particularly interested in the kinematics were the momentum $s$ is
soft, that is, small. In both diagrams the sub-diagram corresponding to
$c\bar c\to gg$ of Fig.~\ref{fig:ccggagain} is much as before. That
is, the internal quark is far off the mass shell, and the two external
gluons are hard. As before we attempt to replace the interaction by a
local $c\bar c gg$ vertex.

\begin{myfig}{Feynman diagrams representing soft gluon radiation as a
correction to the amplitude for $c\bar c$ annihilation of
Fig.~\ref{fig:ccggagain}. The soft gluon carries momentum $s$.} 
\label{fig:softg}
\begin{fmfgraph*}(60,30)
\fmfstraight
\fmfleft{pp,j,i,phantom}
\fmfright{ppp,l,k,r}
\fmfv{label=$s$}{r}
\fmfv{label=$\frac12P+q$}{i}
\fmfv{label=$\frac12P-q$}{j}
\fmfv{label=$\frac12P+k-\frac12s$}{k}
\fmfv{label=$\frac12P-k-\frac12s$}{l}
\fmf{gluon,tension=0}{r,z}
\fmf{quark}{i,z,x}
\fmf{quark,tension=0}{x,y}
\fmf{quark}{y,j}
\fmf{gluon,tension=1}{x,k}
\fmf{gluon,tension=2}{y,l}
\end{fmfgraph*} 

\begin{fmfgraph*}(60,30)
\fmfstraight
\fmfleft{phantom,j,i,pp}
\fmfright{r,l,k,ppp}
\fmfv{label=$s$}{r}
\fmfv{label=$\frac12P+q$}{i}
\fmfv{label=$\frac12P-q$}{j}
\fmfv{label=$\frac12P+k-\frac12s$}{k}
\fmfv{label=$\frac12P-k-\frac12s$}{l}
\fmf{gluon,tension=0}{z,r}
\fmf{quark}{i,x}
\fmf{quark,tension=0}{x,y}
\fmf{quark}{y,z,j}
\fmf{gluon,tension=2}{x,k}
\fmf{gluon,tension=1}{y,l}
\end{fmfgraph*} 
\end{myfig}

Let us analyze the emission of the soft gluon. In the first diagram of
Fig.~\ref{fig:softg} the quark line emitting the gluon gives
\begin{eqnarray*}
\frac{\gamma\cdot(\frac12P+q-s)+m_c}{(\frac12P+q-s)^2-m_c^2}
&\gamma^\mu&
\frac{\gamma\cdot(\frac12P+q)+m_c}{(\frac12P+q)^2-m_c^2}\cr\cr
&\approx&
\frac{(1+\gamma^0)}{-2s^0+s^2/m_c}
~\gamma^\mu~
\frac{(1+\gamma^0)m_c}{(\frac12P+q)^2-m_c^2}\cr\cr
&\to&
\frac{-\gamma^0}{s^0}\left[\left(\frac{1+\gamma^0}2\right)
\frac{m_c}{(\frac12P+q)^2-m_c^2}\right].\cr
\end{eqnarray*}
We have left the propagator on the right side unexpanded since it will
be amputated when we compute an amplitude. Similarly, the second
diagram in Fig.~\ref{fig:softg} gives
\begin{eqnarray*}
\frac{\gamma\cdot(-\frac12P+q)+m_c}{(-\frac12P+q)^2-m_c^2}
&\gamma^\mu&
\frac{\gamma\cdot(-\frac12P+q+s)+m_c}{(-\frac12P+q+s)^2-m_c^2}\cr\cr
&\to&
\left[\frac{m_c}{(\frac12P-q)^2-m_c^2}
\left(\frac{1-\gamma^0}2\right)\right]\frac{+\gamma^0}{s^0}.\cr
\end{eqnarray*}
We see immediately that although each diagram is separately infrared
divergent, \ie, they contain a singularity as $s^0\to0$, the divergence
cancels in the sum. Physically the cancellation can be easily
understood: the coupling involves $\gamma^0$, which corresponds to the
``charge'' of the quark, but $c$ and $\bar c$ have opposite charges.

To see that this physical intuition is correct, let's compute the next
term, involving the current $\gamma^i$. Using
\(\frac12P+q=(m_cc,\vec0)+(\frac1cE,\vec q)$ and neglecting $q^0$ and
$s$,  the Dirac structure of the
soft gluon-quark interaction is
\begin{eqnarray*}
[\gamma\cdot(\frac12P+q-s)+m_c]&\gamma^\mu&[\gamma\cdot(\frac12P+q)+m_c]\cr\cr
&\to&
[m_c(1+\gamma^0)+\gamma\cdot q]
\,\gamma^\mu\,      [m_c(1+\gamma^0)+\gamma\cdot q]\cr\cr
&=& 2m_c^2(1+\gamma^0)\delta^{\mu}_0+m_c[\gamma\cdot q\gamma^\mu(1+\gamma^0) +
(1+\gamma^0)\gamma^\mu\gamma\cdot q]+\cdots\cr\cr
&=&\cases{2m_c^2(1+\gamma^0)-4m_c\vec\gamma\cdot\vec q& for $\mu=0$\cr
-2m_c(1+\gamma^0)q^i&for $\mu=i$\cr}\cr
\end{eqnarray*}
The leading term (for  $\mu=0$) is what we computed above. The new
terms of interest are the ones of order $m_c$. The second diagram
gives, similarly
\begin{eqnarray*}
[\gamma\cdot(-\frac12P+q)+m_c]
&\gamma^\mu&[\gamma\cdot(-\frac12P+q+s)+m_c]\cr\cr
&\to&
\cases{-2m_c^2(1-\gamma^0)+4m_c\vec\gamma\cdot\vec q& for $\mu=0$\cr
-2m_c(1-\gamma^0)q^i&for $\mu=i$\cr}\cr
\end{eqnarray*}
So we see that there is at this order an infrared divergence in these
diagrams. It appears only in the $\mu=i$ terms. However, it does not
afflict decays of $S$-wave charmonium since the factor $q^i$ changes
$L$ by one.\footnote{The amplitude has $\vec q\cdot\vec\epsilon$, where
$\vec \epsilon$ is the gluon polarization vector.} Formally, the
operator with the divergent coefficient is $\chi^\dagger\darr D\psi$,
rather than $\chi^\dagger\psi$. But
\[
\langle0|\chi^\dagger\darr D\psi|\eta_c\rangle=0.
\]
However $L=1$ operators do interpolate for $P$-wave quarkonium.

Exercise: Using the methods of Sec.~\ref{sec:spinsymmetry} show that
\begin{eqnarray*}
\langle0|\chi^\dagger\vec\sigma\cdot\darr D\psi|h_c\rangle &=& 0\cr\cr
\langle0|\chi^\dagger\vec\sigma\cdot\darr D\psi|\chi_1\rangle &=& 0
\end{eqnarray*}
and find the relation between
\(\langle0|\chi^\dagger\vec\sigma\cdot\darr D\psi|\chi_0 \rangle\)
and
\(\langle0|\chi^\dagger\vec\sigma\cdot\darr D\psi|\chi_2 \rangle\).

In the color singlet model there seems to be no way out of this
problem with infrared divergences in the decay of $P$-wave
quarkonium. Let us emphasize that the problem is that we have found an
infrared divergence in the calculation of the matching of a
coefficient, which should be given by short distance physics. Usually
one finds a way to shove the infrared divergences into matrix elements
of operators, but there is no candidate for an operator into which to
shove this infrared divergence.

However, in NRQCD there is a neat resolution to this problem.\cite{BBL} In
leading order, the decay is described by local interactions
\begin{equation}
\label{eq:pwaveLeff}
\delta{\cal L}=
\frac{f_1(^3P_0)}{m^4c^3}{\cal O}_1(^3P_0)
+\frac{f_8(^3S_1)}{m^2c}{\cal O}_8(^3S_1)
\end{equation}
where $f$'s are coefficients determined by matching and ${\cal O}$ are
operators. The subscript denotes the relative color of the quark
bilinears, singlet or octet, and in parenthesis we have  their
${}^JL_S$ number. The operators are
\begin{eqnarray*}
{\cal O}_1(^3P_0) & = & 
\frac13\psi^\dagger(\frac{i}2\darr D\cdot\sigma)\chi
\chi^\dagger(\frac{i}2\darr D\cdot\sigma)\psi\cr\cr
{\cal O}_8(^3S_1) & = & 
\psi^\dagger(\sigma^iT^a)\chi
\chi^\dagger(\sigma^iT^a)\psi
\end{eqnarray*}

\begin{myfig}{Feynman diagram for $c\bar c$ annihilation into light
quarks. It represents the leading contribution to the matching of
the color octet operator~${\cal O}_8(^3S_1)$.}
\label{fig:octetannih}
\begin{fmfgraph}(60,20)
\fmfleft{j,i}
\fmfright{l,k}
\fmf{quark}{i,x,j}
\fmf{quark}{l,y,k}
\fmf{gluon}{x,y}
\end{fmfgraph}
\end{myfig}

Exercise: Show that at tree level the diagram of
Fig.~\ref{fig:ccggagain} gives the matching to the $P$ wave operator
\begin{equation}
\label{eq:matchfone}
{\rm Im}\,f_1(^3P_0)=\frac{3\pi C(R)}{2N_c}\alpha_s^2
\end{equation}
while the diagram of Fig.~\ref{fig:octetannih}, where there are $n_f$
species of light quarks in the possible final state on the right,
gives the $S$-wave coefficient
\begin{equation}
\label{eq:matchfeight}
{\rm Im}\,f_8(^3S_1)=\frac{\pi n_f}6\alpha_s^2
\end{equation}

\subsection{Power Counting}
\label{sec:powercounting}
\subsubsection{Decays of $S$-wave Quarkonium}
For this discussion to make any sense we must show that the two
operators in Eq.~(\ref{eq:pwaveLeff}) contribute to the same (leading)
order in $1/c$.  It is instructive though to begin by revisiting the
decay of the $S$-wave quarkonium. Let us set the normalization
arbitrarily by saying that the contribution from the operator ${\cal
O}_1(^1S_0)$ is of order one:
\[
{\cal O}_1(^1S_0):\qquad\qquad
\langle\eta_c|\psi^\dagger\chi\chi^\dagger\psi|\eta_c\rangle\sim1
\]
Consider other possibilities for $\delta{\cal L}$:
\[
{\cal O}_1(^3S_1):\qquad\qquad
\langle\eta_c|\psi^\dagger\sigma^i\chi\chi^\dagger\sigma^i\psi|\eta_c\rangle
\]
This vanishes by spin-symmetry. However, we can include,
perturbatively,  interactions
that break spin-symmetry and then
the result need not vanish. Recall
\[
{\cal L}=\cdots+\frac{c_Fg}{2mc^{5/2}}\psi^\dagger
\vec\sigma\cdot\vec{\tilde B}\psi.
\]
Now $\vec\sigma\cdot\vec{\tilde B}$ flips spins, but also changes
relative color from singlet to octet. One can restore the quarks into
a singlet color combination by inserting, in addition, an electric
dipole (E1) operator 
\[
{\cal L}=\cdots+\frac{g}{2mc^{3/2}}\psi^\dagger\vec{\tilde A}\cdot
\vec\nabla\psi.
\]
Thus we estimate
\[
\langle\eta_c|{\cal O}_1(^3S_1)|\eta_c\rangle
\sim\left(\frac{g}{c^{5/2}}\frac{g}{c^{3/2}}\right)^2
\sim\left(\frac{g^2}{c}\right)^2\frac1{c^6}.
\]
Consider next
\[
{\cal O}_8(^1S_0):\qquad\qquad
\langle\eta_c|\psi^\dagger T^a\chi\chi^\dagger T^a\psi|\eta_c\rangle.
\]
Inserting the E1 operator twice (once is not enough because it has
$L=1$), we estimate its order to be
\[
\left(\frac{g^2}{c^3}\right)^2\sim\left(\frac{g^2}{c}\right)\frac1{c^4}
\]
The operator ${\cal O}_8(^1P_1)$ requires only one E1 insertion,
but comes built in with a $1/c^2$ suppression:
\[{\cal O}_8(^1P_1):\qquad\qquad
\frac1{m^2c^2}\langle\eta_c|\psi^\dagger T^a\darr D\chi
\chi^\dagger T^a\darr D\psi|\eta_c\rangle
\sim\frac1{c^2}\left(\frac{g}{c^{3/2}}\right)^2\sim
\left(\frac{g^2}{c}\right)\frac1{c^4}.
\]

For these non-perturbative matrix elements\hskip2pt\footnote{The
matrix elements are considered to all orders in the zeroth order
lagrangian of NRQCD. We have used perturbation theory only with
respect to the dimensionful parameter $1/c$.} the strong coupling
constant should be taken large, $g^2/c\sim\alpha_s\sim1$. We conclude
that the corrections to the $\eta_c$ decay width from \( {\cal
O}_8(^1S_0) \) and \( {\cal O}_8(^1P_1) \) are suppressed by $1/c^4$,
and from \( {\cal O}_1(^3S_1) \) by $1/c^6$. The leading correction is
in fact of order $1/c^2$ and arises from an operator with the same
quantum numbers as the leading operator, but with two derivatives:
\[
{\cal P}_1(^1S_0):\qquad\qquad
\frac1{m^2c^2}
\langle\eta_c|\psi^\dagger\darr D^2\chi\chi^\dagger\psi|\eta_c\rangle
\sim\frac1{c^2}
\]
\subsubsection{Decays of $P$-wave Quarkonium}
We now perform an analogous analysis for the case of decays of
$P$-wave states. The operators we focus on are the ones in the
effective lagrangian of Eq.~(\ref{eq:pwaveLeff}). \( {\cal O}_1(^3P_0) \)
has just the right quantum numbers to interpolate the color singlet
$P$-wave states,
\[
\frac1{m^4c^3}\langle P|{\cal O}_1(^3P_0)|P\rangle
\sim\frac1{c^3}
\]
while \( {\cal O}_8(^3S_1) \) requires an E1 transition,
\[
\frac1{m^2c}\langle P|{\cal O}_8(^3S_1)|P\rangle
\sim\frac1{c}\left(\frac{g}{c^{3/2}}\right)^2\sim
\left(\frac{g^2}{c}\right)\frac1{c^3}.
\]
We see that these two operators contribute to the $P$-wave decay width
at the same order in the $1/c$ expansion. It is easy to see that there
are no other operators that contribute at this order.

We can now use the Wigner-Eckart theorem to express the decay widths
in terms of reduced matrix elements
\begin{eqnarray*}
\langle{\cal O}_1\rangle & \equiv & 
\frac{|\langle0|\chi^\dagger(\frac{i}2\darr D\cdot\vec\sigma)\psi
                                |\chi_{c0}\rangle|^2}{2M_\chi}\cr\cr
\langle{\cal O}_8\rangle & \equiv & 
\frac{\langle\chi_{c0}|\psi^\dagger\sigma^iT^a\chi\chi^\dagger\sigma^iT^a\psi
                                |\chi_{c0}\rangle}{2M_\chi}
\end{eqnarray*}
Using the results of matching at leading order in
Eqs.~(\ref{eq:matchfone}) and~(\ref{eq:matchfeight}) the widths for
the four $P$ wave particles are given in terms of these two reduced
matrix elements:
\begin{eqnarray*}
\Gamma(h_c) & = & \frac{5\pi\alpha_s^2}{6m_c^2}\langle{\cal
O}_8\rangle \cr\cr
\Gamma(\chi_{c0}) & = & \frac{4\pi\alpha_s^2}{m_c^4}\langle{\cal
O}_1\rangle + \frac{n_f\pi\alpha_s^2}{3m_c^2}\langle{\cal
O}_8\rangle \cr\cr
\Gamma(\chi_{c1}) & = & \frac{n_f\pi\alpha_s^2}{3m_c^2}\langle{\cal
O}_8\rangle \cr\cr
\Gamma(\chi_{c2}) & = & \frac{16\pi\alpha_s^2}{45m_c^4}\langle{\cal
O}_1\rangle + \frac{n_f\pi\alpha_s^2}{3m_c^2}\langle{\cal
O}_8\rangle 
\end{eqnarray*}

We are now in a position to understand the cancellation of infrared
divergences. In the expression
\[
\Gamma(\chi_{c0})  =  \frac{{\rm Im}\,f_1(^3P_0)}{m^4c^3}\langle{\cal
O}_1\rangle + \frac{{\rm Im}\,f_8(^3S_1)}{m^2c}\langle{\cal
O}_8\rangle 
\]
the infrared divergence in ${\rm Im}\,f_1(^3P_0)$ is precisely
cancelled by a divergence in the matrix element $\langle{\cal
O}_8\rangle $. The resolution of this problem is responsible for much
of the renewed interest in NRQCD.

\subsection{Decays: Summary}
In computing decays of quarkonium into light hadrons we include in the
effective lagrangian local four-fermion operators
\[
\delta{\cal L} = \sum_n f_n{\cal O}_n
\]
and compute
\[
\frac12\Gamma=\sum_n{\rm Im}\,f_n \langle{\cal O}_n\rangle, 
\]
where \( {\rm Im}\,f_n \) is computed by matching to QCD. The relative
importance of these terms is determined by power counting in $1/c$, 
using the minimal number of required interactions from the multipole
expanded  NRQCD effective lagrangian, {\it and using} $g^2/c\sim\alpha\sim1$.

The operators ${\cal O}_n$ are of the form
\[
\psi^\dagger\kappa_m\chi\chi^\dagger\kappa_n\psi
\]
with $\kappa_m$ a tensor product of $\darr D$, $T^a$ and $\sigma^i$.

\end{fmffile}

\section{Concluding Remarks}
There are many open questions that remain un-addressed. In the standard
presentation of the subject~\cite{BBL} the color octet bilinears in
the lagrangian of NRQCD, such as ${\cal O}_8(^3S_1)$ of
Eq.~(\ref{eq:pwaveLeff}), are interpreted as literally producing and
annihilating color octet states. It is not quite clear what these
states are. It is reasonable to assume they correspond to the hybrid
states of Sec.~\ref{sec:hybrids}. But doing so would invalidate the
standard analysis for power counting,\cite{BBL} which requires the
size of this states to still be of the order of $a_{\rm Bohr}$ rather
than the bag radius $\LQCD^{-1}$. 

A closely related question is whether the non-perturbative potential,
which grows linearly with distance at large distances, should be used
in NRQCD. The Born-Oppenheimer approximation certainly suggests this
is the right approach. However, it remains possible that the
condensates of the higher dimension operators fully incorporate the
non-perturbative information relevant to all the questions that NRQCD
attempts to answer. In that case, including both the non-perturbative
potential and the condensates of higher dimension operators would
result in double-counting.

Another question that remains unsolved is the rather arbitrary
assumption that $g^2/c$ must be taken as of order unity in order for
power counting to work, as presented in
Sec.~\ref{sec:powercounting}. It is not enough to insist that $g^2\sim
c$ in general, for this must not be the case in calculations of
matching coefficients of the operators in $\delta{\cal L}$. The
special rule $g^2/c\sim1$ applies only when calculating matrix
elements of operators.

Because of time and space limitations I have not included in these
lectures a discussion of the analysis of quarkonium production in
NRQCD. Yet most of the ongoing work on the subject is in this
realm. There are many good reviews the reader may consult.\cite{Beneke} 

\section*{Acknowledgments}
I would like to thank the organizers of the first Winter School
on "Recent Developements in Hadron Physics" 
at the  Asia Pacific Center for Theoretical Physics for their
invitation to participate in this school, their financial assistance
and, most notably, for their warm hospitality. This work was supported
in part by the Department of Energy under grant DOE-FG03-97ER40546.

\eject

\end{document}